\documentclass[12pt]{iopart}

\usepackage{iopams} 
\usepackage[all]{xy}
\usepackage{graphicx}
\usepackage{subfigure}
\newcommand{\R}{\mathbb{R}}

\newcommand{\T}{\mathbb{T}}
\newcommand{\Q}{\mathbb{Q}}
\newcommand{\A}{\mathbb{A}} 
\begin{document}
\title[Electrons in deterministic quasicrystalline potentials]{Electrons in deterministic quasicrystalline potentials and hidden conserved quantities}
\author{P~Kalugin$^1$ and A~Katz$^2$}
\address{$^1$ Laboratoire de Physique des Solides, UMR CNRS 8502, Bât 510, Universit\'e Paris Sud, 91405 Orsay cedex, France}
\address{$^2$ Centre de Physique Th\'eorique, Ecole Polytechnique, CNRS, 91128
Palaiseau, France}
\ead{kalugin@lps.u-psud.fr}
\begin{abstract}
We propose an ansatz for the wave function of a non-interacting quantum particle in a deterministic quasicrystalline potential. It is applicable to both continuous and discrete models and includes Sutherland's hierarchical wave function as a special case. The ansatz is parameterized by a first cohomology class of the hull of the structure. The structure of the ansatz and the values of its parameters are preserved by the time evolution. Numerical results suggest that the ground states of the standard vertex models on Ammann-Beenker and Penrose tilings belong to this class of functions. This property remains valid for the models perturbed within their MLD class, e.g. by adding links along diagonals of rhombi. The convergence of the numerical simulations in a finite patch of the tiling critically depends on the boundary conditions, and can be significantly improved when the choice of the latter respects the structure of the ansatz.
\end{abstract}
\pacs{71.23.Ft}
\submitto{\JPA}
\section{Introduction}
More than thirty years after the discovery of quasicrystals a satisfactory quantum theory of quasicrystalline solids is still far from being constructed. Even for the simplified model of non-interacting electrons in quasicrystalline potentials, nothing is known which would look like as a generalization of the band theory for crystals. Moreover, there is a striking difference between the one-dimensional case, where considerable progress has been achieved and the more physical case of higher dimensions, where the results are quite scant (see a survey \cite{damanik2012spectral}). After an initial enthusiasm, the problem was nearly abandoned, despite its obvious importance for the physics of quasicrystals. The purpose of the present paper is to reopen this question, while taking into account recent results in mathematical studies of aperiodic structures.
\par 
Let us consider a single-particle Schr\"odinger Hamiltonian:
\begin{equation}
\label{sh_operator}
H=-\Delta +U(\mathbf{x}).
\end{equation}
The main rigorous results for the operator (\ref{sh_operator}) with a quasicrystalline potential $U(\mathbf{x})$ in arbitrary dimensions, namely the theorems on labelling of the spectral gaps \cite{bellissard1992gap} and on existence of the integrated density of states \cite{lenz2005ergodic}, are obtained by the methods of operator algebra. However, the solid state physics traditionally deals with the particle {\em states}. In particular, Bloch wave functions have became {\em lingua franca} of the quantum theory of crystals. Their application goes far beyond the single-particle problem in a perfect crystal, since Bloch states are used as elementary bricks in perturbative approaches to much more complex multi-body problems. One of the reasons for the success of Bloch states is their universal character. In fact, the definition of the quasimomentum depends only on the symmetry of the crystal and not on the details of the structure within a unit cell. No such universality is known for quasicrystals.
\par
To gain greater insight into why Bloch states are not suited for quasicrystals, let us recall some facts about the generalized eigenstates of (\ref{sh_operator}) for one-dimensional quasiperiodic potentials. One of the early results in the field \cite{dinaburg1975one} suggests that under some rather restrictive conditions on $U$ the solutions of the stationary Schr\"odinger equation $H\psi=\mathcal{E}\psi$ have the form $\Psi_0 \exp(\rmi \mathbf{k}\mathbf{x})$ with quasiperiodic $\Psi_0$, which can be considered as a generalization of Bloch states to the quasiperiodic setting. The conditions for that includes the smallness of the potential with respect to $\mathcal{E}$ and the rapid decrease of its Fourier coefficients. However, it soon became clear that neither condition can be dropped. In fact, it has been shown that the generalized eigenstates of the almost Mathieu operator become localized when the strength of the potential exceeds a critical value (see \cite{last1994almost} for a review). Similarly, the study of the Fibonacci chain\footnote{In mathematical literature the term {\em quasiperiodic function} is often used in the restricted sense of Bohr almost periodic function \cite{bohr1925theorie} with finitely generated frequency module. Although neither the Fibonacci chain nor the potentials arising in the study of quasicrystals qualify for this definition, in physical literature both are called quasiperiodic.} \cite{kohmoto1987critical} reveals that the eigenfunctions for this model are always unbounded and therefore cannot have the form suggested in \cite{dinaburg1975one}. On the other hand, the ``quasi-Bloch'' eigenstates of the model \cite{dinaburg1975one} can be understood perturbatively, as a superposition of functions $\exp\left(\rmi (\mathbf{k}+\mathbf{k}_m)\mathbf{x}\right)$ where $\mathbf{k}_m$ are the wave vectors of the harmonics of $U$. A similar formal expression can be written for the Fibonacci chain, however, the results \cite{kohmoto1987critical} suggest that the corresponding series diverges. In other words, multiple scattering by the harmonics of Fibonacci potential leads to delocalization of the wave function in the momentum space. The crucial role in such delocalization is played by slowly decreasing amplitudes of Fourier harmonics of $U$ in the Fibonacci chain. In this regard, quasicrystalline potentials have a similar asymptotic behaviour of their Fourier spectrum. Therefore electrons in quasicrystals also should exhibit delocalization in the momentum space \cite{kitaev1988electronic} and thus should not admit ``quasi-Bloch'' eigenstates.  
\par
Since solving the eigenstate problem in quasicrystalline potentials proves to be a challenging task, it makes sense to approach it by steps. Solving the problem entirely would amount to construct for (\ref{sh_operator}) a complete system of generalized eigenstates $\psi_\Lambda$ (the completeness here is understood in the sense of nuclear spectral theorem \cite{dubin1990quantum}). The spectral parameter $\Lambda$ would thus represent a full set of ``good quantum numbers''. However, if only a subset of good quantum numbers is known, one can still use them to parameterize the solution of the time-dependent Schr\"odinger equation:
\begin{equation}
\label{schroedinger}
\rmi \dot \psi_\Lambda(\mathbf{x},t)=(-\Delta +U(\mathbf{x})) \psi_\Lambda(\mathbf{x},t).
\end{equation}
In this setting, $\psi_\Lambda$ is no more a generalized  eigenstate, although the parameter $\Lambda$ represents a conserved quantity. Finding such parameterization would indicate that the motion of a quantum particle in quasicrystals is at least partially integrable. This is the approach we follow in this paper.
\par
The paper is organized as follows. In Section \ref{sec:qc_hulls} we recall the construction of the hull of the structure in the general case. Then we constrain our scope to the case of structures characterized by quadratic irrationalities and compatible with matching rules, in which case the hull can be approximated by a CW-complex. In Section \ref{sec:ansatz} we propose a formal ansatz for the wave function $\psi_\Lambda$ of (\ref{schroedinger}) and analyze the uniqueness of its parameterization. In Section \ref{sec:tight-binding} we generalize the ansatz to the case of tight-binding models of quasicrystals. We show then that the parameters of the ansatz are conserved through the time evolution of the wave function. Section \ref{sec:numerical} is devoted to the numerical study of the problem in the discrete setting. First we propose special boundary conditions respecting the structure of the ansatz for the stationary Schr\"odinger equation on a finite patch of a quasiperiodic tiling.   Then we report numerical results suggesting that the ground states in various classes of tight-binding models actually belong to the proposed class of solutions.
\section{Quasicrystals and their hulls}\label{sec:qc_hulls}
The structures of both crystals and quasicrystals are highly repetitive in space. Whereas crystals are fundamentally so because of the perfect translational symmetry of the lattice, the repetitivity of quasicrystals is only approximate. The ambiguous concept of approximate translational invariance finds a rigorous expression in terms of the continuous hull of the structure. 
\par
Historically, the notion of the hull originated in the study of almost periodic (a.p.) functions. The hull of an a.p. function is defined as the closure of its orbit under translations in an appropriate topology. The latter is chosen accordingly to the kind of a.p. functions considered (see \cite{lagarias2000mathematical} for a review). For instance for the Bohr a.p. functions, the natural choice is the $L^\infty\mbox{-norm}$ topology \cite{jitomirskaya1994operators}, whereas for the Besicovitch a.p. functions the compact open topology is generally used. Bellissard \cite{bellissard1986k} extended the notion of the hull to operator algebras. Namely, for an operator $A \in \mathcal{B}\left(L^2(\R^d)\right)$ its hull is defined as the closure of the orbit of its translates (that is, of the set $\{U_{\mathbf{x}}AU_{\mathbf{x}}^{-1}, \, \mathbf{x} \in \R^d\}$, where $U_{\mathbf{x}}$ is the unitary operator corresponding to the translation by the vector $\mathbf{x}$) in the strong operator topology of $\mathcal{B}\left(L^2(\R^d)\right)$. The hull can also be defined for an unbounded operator such as (\ref{sh_operator}) as the closure of the orbit of its resolvent. For the one-particle Schr\"odinger operator (\ref{sh_operator}) the resulting space is homeomorphic to the hull of the potential $U$ defined as the $L^\infty\mbox{-weak}$ closure of its orbit under translations \cite{bellissard1992gap}. Note that this definition does not require the almost periodicity of $U$; in fact is applicable to any measurable essentially bounded function.
\par
The construction of the continuous hull according to \cite{bellissard1992gap,bellissard1986k} yields a compact metrizable space $\Omega$ together with the action of the translations on it. Alternatively, the metric topology on the hull can be defined directly by its construction. In this approach, one starts with a metric of ``approximate match'' in the physical space. In this metric, the ``distance'' $D(\mathbf{x}, \mathbf{y})$ between the points $\mathbf{x}$ and $\mathbf{y}$ is small when the patterns surrounding $\mathbf{x}$ and $\mathbf{y}$ almost coincide within a large area. More specifically, for the case of structures characterized by a uniformly continuous (pseudo) potential $U$ one can set $D(\mathbf{x}, \mathbf{y})$ as the lower bound of all real $\epsilon$ for which $|\mathbf{r}|<\epsilon^{-1}$ implies $|U(\mathbf{x}+\mathbf{r}) - U(\mathbf{y}+\mathbf{r})|<\epsilon$.
The definition of $D(\mathbf{x}, \mathbf{y})$ can be adapted to other models of atomic structures. For instance, if the latter is modelled by discrete point sets representing atomic positions, one can use Hausdorff distance to measure the difference between the patches of the structure surrounding $\mathbf{x}$ and $\mathbf{y}$ \cite{radin1992space}. In all cases, one endows the $d\mbox{-dimensional}$ physical space $E$ with the metric $D(\mathbf{x}, \mathbf{y})$ (which is very different from the ``natural'' euclidean metric of $E$!). The completion of $E$ with respect to $D(\mathbf{x}, \mathbf{y})$ yields the continuous hull of the structure.
\par
The hull can be constructed for any structure, but in the general case one obtains a fairly complicated space. Notorious exceptions are crystals and quasicrystals. For crystals, the continuous hull is a torus $\T^d$, corresponding to the unit cell, while for the quasicrystals it is ``almost a torus''. More precisely, let us consider the dynamical system $(\Omega, \R^d)$, defined by the natural action of the translations of the physical space on $\Omega$. Then the maximal equicontinuous factor \cite{math2012maximal} of $(\Omega, \R^d)$ in the case of quasicrystals is always a torus $\T^N$ of dimension  $N>d$ (note that in the general case this factor is a pro-torus \cite{hofmann2006structure}). The construction of the maximal equicontinuous factor yields also a natural surjective map $\pi: \Omega \to \T^N$, called in \cite{kellendonk2013topological} {\em the torus parameterization}. Since the quasicrystals have pure point diffraction spectrum, the spectrum of the dynamical system $(\Omega, \R^d)$ is also pure point \cite{baake2004dynamical} and the map $\pi$ is one-to-one almost everywhere in the translation-invariant measure on $\Omega$ \cite{barge2013,math2012maximal}.
\par
Let us clarify the relation between the construction above and the traditional ``cut-and-project'' modelling of quasicrystalline structures. Since the parallel translations of physical space act naturally on $\Omega$, fixing the origin in $E$ with the corresponding point $x_0 \in \Omega$ defines a map $\mu$ from $E$ to the orbit of this point:
$$
\mu: \mathbf{x} \mapsto x_0 + \mathbf{x}
$$
(to keep the notation simple, we shall denote the action of $\textbf{x}$ on $x_0$ by $x_0 + \mathbf{x}$). The cut-and-project approach ignores $\Omega$ and deals with the irrational winding of $E$ on $\T^N$ given by the composite map $\pi \circ \mu$ only:
\begin{equation}
\label{hull_and-torus}
\xymatrix{E\ar[r]^{\mu} &\Omega \ar[r]^{\pi}& \T^N}.
\end{equation}
\par
One should emphasize the importance of the middle term $\Omega$ in (\ref{hull_and-torus}). Indeed, while in the cut-and-project approach the distribution of the potential (or the electron density or any other characteristic of the structure for that matter) in $E$ is obtained as a pullback of some function $u$ on $\T^N$ by $\pi \circ \mu$, the degree of regularity of $u$ is unclear. One the one hand, requiring the continuity of $u$ on $\T^N$ would be too restrictive. In fact, this is valid for modulated crystalline structures only \cite{janner1980symmetry}. On the other hand one cannot obviously abandon the requirement of continuity altogether. Actually, the very notion of ``atomic surface'' widely used in the structure analysis of quasicrystals \cite{bak1986icosahedral} arises as an attempt to impose some sort of piecewise continuity of $u$ in the direction transverse to that of $E$ on $\T^N$. Although this approach gives satisfactory results in X-ray structure determination \cite{cornier1991neutron}, it may only be an approximation. In fact, it is clear on physical grounds that the precise location of any atom depends on the location of its neighbours; as is well known, the atomic surface is dissected in the so called ``existence domains'' of local structures, each of which with its own position. Even if the influence of neighbours falls off quickly with the distance, this is incompatible with the piecewise continuity of the atomic surface, or of any $u$ on $\T^N$.
\par
The notion of the hull resolves all these difficulties in an aesthetically appealing way. In fact, since the function $U(\mathbf{x})$ is continuous in the topology of the metric $D(\mathbf{x}, \mathbf{y})$, it can be extended to a continuous function on $\Omega$. Therefore, there exists $u \in C(\Omega)$ such that $U(\mathbf{x})$  is a pullback of $u$ by $\mu$:
\begin{equation}
\label{pullback_u}
U(\mathbf{x})=u\left(\mu(\mathbf{x})\right).
\end{equation} 
In other words, the topology of $\Omega$ is, by construction, just as strong as it is needed to make $u$ continuous.
\par
Up to now we considered the atomic structure as primary data, and the hull $\Omega$ was constructed subsequently from it. However, as evidenced by the case of crystals, the resulting space may be much less variable that the structure itself. As we shall see below, the considerations of structural stability lead to similar conclusions for the hull of quasicrystals. This makes reasonable to shift the standpoint and interpret the hull (together with the map $\mu:E \to \Omega$) as a host for many possible structures. Therefore, the hull will play the role of a natural  framework for the structure, for instance, the only admissible potentials $U$ will be those obtained as pullbacks of continuous functions on $\Omega$. We are convinced of the fruitfulness of this point of view on the role of the hull, which is currently emerging among the specialists (see \cite{baake2013aperiodic} for a comprehensive review). The ultimate expression of it would be considering the hull (together with the action of translations on it) as an embodiment of the symmetry of the quasicrystal, not only in studying the Schr\"odinger equation, but also in other aspects, such as the structure analysis. Note that the compactness of $\Omega$ fits perfectly this scheme. Indeed, since a continuous function on a compact can be uniformly approximated with any desired accuracy by functions on finite sets, the modelling  of quasicrystalline structures is not much different from the case of crystals, where once the symmetry class is fixed, the structure is determined by a finite number of atomic positions. 
\par
In what follows we shall often deal with functions on $\Omega$ with different degrees of the regularity. To measure this degree, one needs something more than just the topology of $\Omega$. The hull is already equipped with two additional structures, namely the dynamical system on it associated with the action of the translations of the physical space and the metric inherited from $D(\mathbf{x}, \mathbf{y})$. An example of using the first structure is provided by the so called weakly patterns equivariant (PE) functions \cite{kellendonk2003pattern}. These functions are originally defined on $E$, but by construction they correspond to pullbacks of continuous functions on $\Omega$. More specifically, the space of weakly PE-functions is a closure (in an appropriate Fr\'echet topology) of the transversally locally constant functions on $\Omega$ having a $C^\infty$ pullback on $E$. By construction, the weakly PE-functions are $C^\infty$ in the tangential direction, but transversally they are merely continuous. In this paper, however, we are mostly interested in measuring the regularity in the transversal direction. Such measure can be provided through the metric space structure of $\Omega$, for instance, one may consider functions satisfying H\"older conditions with various exponents. Instead of H\"older norm one could use any other norm tailored to inhibit strong variations on small distances. An example of this approach is given in Section \ref{sec:tight-binding} where we introduce a family of weighted Hilbert norms on the canonical transversal of the hull.
\par
So far we considered quasicrystals from the purely geometrical point of view. However, as real physical systems, these materials should be stabilized by short-range interactions between atoms. These considerations impose further constraints on $\Omega$, more specifically on its singular subset (that is a subset containing precisely those points for which the map $\pi$ in (\ref{hull_and-torus}) is not injective). This can be understood from the following reasoning. Let $x_1$ and $x_2$ be two different points of $\Omega$ such that $\pi(x_1)=\pi(x_2)$. The latter clearly holds also when $x_1$ and $x_2$ are translated by any $x \in \R^d$. Then the function $\Delta(x)=D(x_1+x, x_2+x)$ characterizes what is called in the physical literature the response of the structure to an infinitesimal phason shift. The points $x_1$ and $x_2$ are proximal for the dynamical system $(\Omega, \R^d)$  \cite{barge2013}, that is $\inf\left(\Delta(x)\right)=0$. However, the stability of quasicrystal requires unboundedness and connectedness of the set $\{x,\;\Delta(x)>\delta\} \subset \R^d$ for some $\delta>0$ since otherwise no local interaction could enforce a globally coherent choice between the patterns corresponding to $x_1$ and $x_2$. In other words, the propagation of the quasiperiodic order requires that the regions of the structure exhibiting significant rearrangement under an infinitesimal phason shift form a globally connected net.
\par
For the case of tiling-based structure models, the above considerations are usually formulated as the problem of {\em local rules} \cite{levitov1988local, katz1988theory} (in the case of decorated tilings the term {\em matching rules} is also used \cite{katz1995matching}). The global net of rearrangements of tiles under an infinitesimal phason shift in this case consists of $(d-1)\mbox{-dimensional}$ strips, commonly called ``worms'' \cite{socolar1986phonons}. The condition of continuity of the ``worms'' was studied in \cite{katz1995matching,levitov1988local} for the specific case of canonical projection tilings. It was shown that it leads to a constraint on the slope of the irrational winding $\pi \circ \mu$ of (\ref{hull_and-torus}). Namely, in the case of two-dimensional tiling, the existence of strong local rules is only possible when the slope is given by a quadratic irrationality. More precisely, there should exist a quadratic irrational number $a$ such that $E$ meets the points of $\T^N$ with coordinates from $\Q[a]$ on a dense set. Reciprocal results were obtained in \cite{thang1995local}, where it was shown that any canonical projection tiling satisfying the above condition on the irrationality admits strong local rules (for any dimension of $E$). The importance of this condition is strongly corroborated by the fact that so far only quasicrystals with quadratic irrationalities have been discovered in nature. 
\par
The existence of matching rules imposes further constraints on $\Omega$. First of all, the quadratic irrationality condition allows for a natural choice of the so-called transversal space $E_\bot$ (here we consider $E$ as a subspace of $\R^N$, the universal covering of $\T^N$ from (\ref{hull_and-torus})). Namely, one can construct $E_\bot$ as a Galois dual to $E$, considered as a $d\mbox{-dimensional}$ space over $\Q[a]$ (see \cite{thang1995local} for further details). Moreover, since $E+E_\bot$ is a rational subspace of $\R^N$, one can assume without loss of generality that $N=2d$. The duality between $E$ and $E_\bot$ carries through to the singular subspaces of $E_\bot$, which are dual to the hyperplanes of $E$ corresponding to the ``worms''. In the case of the structures described by the so called {\em model sets} (\cite{meyer1972algebraic}, see also \cite{moody2000model} for a survey), this requirement leads to a constraint on the boundary of the acceptance domain (or the ``window''), known as the rationality condition  \cite{gahler2013integral}. This condition is quite a strong one, since, as it is shown in \cite{kalugin2005cohomology}, it implies that $\Omega$ is homeomorphic to the inverse limit of a sequence of CW-spaces $X_n$ with cellular maps $\iota_m: X_{m+1} \to X_m$. More specifically, the construction is based on an arrangement of affine subtori of $\T^N$ of codimension 2, called $\A$ in \cite{gahler2013integral}. The singular cuts are precisely those for which the winding (\ref{hull_and-torus}) of $E$ meets $\A$. Given a sequence of balls $B_n \subset E$ of increasing radius one defines a ``thickened'' arrangements $A_n= \A+B_n$. The spaces $X_n$ are then constructed as a completion of  $\T^N \backslash A_n$ in its inner metric, inherited from the natural euclidean metric of $\T^N$ \cite{kalugin2005cohomology}. The overall situation is described by the following commutative diagram: 
\begin{equation}
\label{lim_hull}
\xymatrix{\Omega\ar@/_1pc/[dd]_\pi \ar[rd]^{\xi_n}\ar[rrd]^{\xi_{n-1}}\\
\dots \ar[r]^{\iota_{n}} & 
X_n \ar[r]^{\iota_{n-1}} \ar[ld]_{\alpha_n} & 
X_{n-1} \ar[r]^{\iota_{n-2}} \ar[lld]_{\alpha_{n-1}} & \dots \\
 \T^N}
\end{equation}
Here the maps $\alpha_m$ arise naturally by construction of $X_m$, and the maps $\xi_m$ exist because of the universal property of the inverse limit. As shown in \cite{kalugin2005cohomology}, there exists a positive integer $n$ such that for all $m \ge n$ the maps $\iota_m$  in (\ref{lim_hull}) are homotopy equivalences (actually, according to \cite{gahler2013integral} the spaces $\Omega$ and $X_n$ are themselves equivalent as objects in the shape category, see \cite{clark2012tiling}). We shall call such space $X_n$ the CW-approximation of the continuous hull $\Omega$.
\par
In what follows we shall always assume that the slope of the winding of $E$ on $\T^N$ is given by a quadratic irrationality and that $\Omega$ fits the diagram (\ref{lim_hull}).
\section{An ansatz for the wave function}\label{sec:ansatz}
In this section we introduce an ansatz for the continuous Schr\"odinger equation (\ref{schroedinger}) with a quasicrystalline potential. Since crystalline potentials are degenerate cases of quasicrystalline ones, they are a good starting point for the formulation of the ansatz. In the case of a periodic potential $U$ the map $\pi$ in (\ref{hull_and-torus}) is an isomorphism. The hull itself is therefore an $N\mbox{-dimensional}$ torus and the map $\mu$ in (\ref{hull_and-torus}) is the universal covering of it ($N$ is thus equal to the dimension of the physical space). Let us consider the Bloch wave function for the quasimomentum $\mathbf{k}$:
\begin{equation}
\label{bloch}
\psi_\mathbf{k}(\mathbf{x},t)=\Psi_0(\mathbf{x},t)\exp(\rmi \mathbf{k}\mathbf{x}),
\end{equation}
where $\Psi_0(\mathbf{x},t)$ is periodic with respect to the crystal lattice. Note that the particular form of the phase multiplier $\exp(\rmi \mathbf{k}\mathbf{x})$ is of no importance since one can always rewrite $\psi_\mathbf{k}(\mathbf{x})$ as
\begin{equation}
\label{bloch2}
\psi_\mathbf{k}(\mathbf{x},t)=\Psi_0(\mathbf{x},t)\exp\left(\rmi \phi(\mathbf{x})\right)
\end{equation}
with some different periodic prefactor $\Psi_0$, as long as $\nabla \phi(\mathbf{x})$ is also periodic and $\phi(\mathbf{x}+\mathbf{d})-\phi(\mathbf{x})=\mathbf{k}\mathbf{d}$ for any lattice translation $\mathbf{d}$. In other words, $\nabla \phi(\mathbf{x})$ is a closed differential 1-form on $\T^N$, and the quasimomentum $\mathbf{k}$ is entirely determined by the de Rham cohomology class of this form.
\par
The expression (\ref{bloch2}) is not immediately generalizable to the case of quasicrystals since their hull is not a manifold and it is not possible to define the de Rham complex directly on $\Omega$. One could circumvent this difficulty by working with differential forms directly in the physical space $E$ and imposing an additional requirement of {\em pattern equivariance}. This approach, yielding to the notion of the pattern equivariant (PE) cohomologies of $E$ \cite{kellendonk2003pattern, kellendonk2006ruelle, boulmezaoud2010comparing}, is too general for our purposes. Instead, we shall use the fact that the \v{C}ech cohomology groups of $\Omega$ and its CW-approximation $X_n$ are isomorphic \cite{kalugin2005cohomology}:
$$
\check{H}^*(\Omega)=\check{H}^*(X_n).
$$ 
Let $\Lambda \in \check{H}^1(\Omega)$ be a class of the first \v{C}ech cohomology group of $\Omega$ with complex coefficients. Consider a CW-approximation $X_n$ of $\Omega$. Since $X_n$ is also a differential manifold with boundary (as a completion of $\T^N \backslash A_n$), one can interpret $\Lambda$ as a class of de Rham cohomology of $X_n$. Let $\omega$ be a closed 1-form on $X_n$ belonging to the class $\Lambda$. Its pullback $\mu^*(\omega)$ is a closed 1-form on the contractible space $E$ and therefore
\begin{equation}
\label{df}
\mu^*(\omega)=\rmd f_{\Lambda}
\end{equation}
for some $C^\infty$ complex valued function $f_{\Lambda}$ on $E$. Consider the following formal ansatz for the wave function:
\begin{equation}
\label{ansatz}
\psi_\Lambda(\mathbf{x},t)=\Psi_0\left(\mu(\mathbf{x}), t \right) \exp\left(2 \pi \rmi f_{\Lambda}(\mathbf{x})\right),
\end{equation} 
where $\Psi_0(y,t)$ is a time-dependent function of $y \in \Omega$, for which we do not assume any regularity for the moment. One remarks the similarity of this expression with (\ref{bloch2}), in particular in that one can freely choose a representative $\omega$ of the class $\Lambda$, since the difference can be absorbed in the pre-exponential factor $\Psi_0$. The Schr\"odinger equation (\ref{schroedinger}) for $\psi_\Lambda(\mathbf{x},t)$ with $U(\mathbf{x})$ of the form (\ref{pullback_u}) then factors formally to:
\begin{equation}
\label{psi0}
-\rmi \dot \Psi_0= \Delta_\| \Psi_0 + 
g_1 \cdot (\nabla_\| \Psi_0)
+ g_2\Psi_0,
\end{equation}
where $\nabla_\|$ and $\Delta_\|$ stand for the gradient and the Laplacian on $\Omega$ in the direction of the physical space and $g_1$ and $g_2$ are defined on $\mu(E)$ by the following equations:
\begin{equation}
\label{g1}
g_1\left(\mu(\mathbf{x})\right)=4 \pi \rmi \nabla f_{\Lambda}(\mathbf{x})
\end{equation}
\begin{equation}
\label{g2}
g_2\left(\mu(\mathbf{x})\right)=-4 \pi^2\left(\nabla f_{\Lambda}(\mathbf{x})\right)^2 +
2 \pi \rmi \Delta f_{\Lambda}(\mathbf{x}) - u\left(\mu(\mathbf{x})\right).
\end{equation}
\par
Here is the pivotal point in the construction of the ansatz.
One can remark that $f_{\Lambda}$ appears in the right-hand side of (\ref{g1}) and (\ref{g2}) exclusively as a derivative, henceforth both expressions are weakly pattern-equivariant. Therefore $g_1$ and $g_2$ can be extended to the entire hull, yielding continuous vector-valued and scalar-valued functions on $\Omega$ respectively. Let us consider now the equation (\ref{psi0}) as a Cauchy problem for $\Psi_0(y,t)$ and set as an initial condition at $t=0$ a tangentially smooth \cite{moore2006global} function from $C(\Omega)$. Then the question arises: will $\Psi_0$ remain continuous for all $t>0$? Although this seems plausible since the right-hand side of (\ref{psi0}) is continuous on $\Omega$, the answer to this question requires a careful analysis of regularity of $\Psi_0$, both in tangential and in perpendicular direction. This quite involved task does not enter into the scope of the present article. We believe however, that the corresponding difficulties are not specific for quasicrystalline potentials and are entirely due to the unboundedness of the Laplacian operator in (\ref{schroedinger}). Indeed, as we shall see in Section \ref{sec:tight-binding}, for discrete models the answer to the above question is positive.
\par
Consider now the spatial behaviour of $\psi_\Lambda$, namely its growth rate. Since $\Psi_0$ is bounded (as a continuous function on the compact space $\Omega$), the growth of $\psi_\Lambda$ depends entirely on that of $f_{\Lambda}$. Recall that (\ref{df}) defines $f_{\Lambda}$ as the integral of a pullback of a closed form $\omega$ on $X_n$, and that $\Lambda$ is the first cohomology class of $\omega$. Let us consider the structure of $H^1(X_n)$ in more details. Since the pullback $\alpha_n^*: H^1(\T^N) \to H^1(X_n)$ is a monomorphism \cite{kalugin2005cohomology}, we can consider the following short exact sequence:
\begin{equation}
\label{split}
\xymatrix{H^1(\T^N) \ar[r]^{\alpha_n^*} & H^1(X_n) \ar[r] & \mathrm{coker}(\alpha_n^*)}.
\end{equation}
This sequence is naturally split, as can be seen from the following arguments. Since the interior of $X_n$ is diffeomorphic to $\T^N\backslash A_n$, every closed 1-form $\omega$ on $X_n$ defines a 1-form on $\T^N\backslash A_n$. Let us denote the average of this form with respect to the Haar measure on $\T^N$ by $\overline{\omega}$. The mapping $\omega \mapsto \overline{\omega}$ factors to the cohomology yielding the map $\epsilon: H^1(X_n) \to H^1(\T^N)$ such that $\epsilon \circ \alpha_n^* = \mathrm{id}$. Hence the sequence (\ref{split}) is left split. It is also right split by the map $\mathrm{coker}(\alpha_n^*) \to H^1(X_n)$ associating each class from $\mathrm{coker}(\alpha_n^*)$ to its representative in $H^1(X_n)$ that vanishes on all 1-cycles of $\T^N$. Therefore one has
\begin{equation}
\label{split_cohom}
H^1(X_n) = H^1(\T^N) \oplus \mathrm{coker}(\alpha_n^*).
\end{equation}
The class $\Lambda \in H^1(X_n)$ from (\ref{ansatz}) is decomposed accordingly into the following sum (since there is a natural isomorphism between $\check{H}^1(\Omega)$ and $H^1(X_n)$ we shall no longer make any distinction between these groups):
$$
\Lambda=  \Lambda_{\rm B} + \Lambda_{\rm S},
$$
where $\Lambda_{\rm B} \in H^1(\T^N)$ and $\Lambda_{\rm S} \in \mathrm{coker}(\alpha_n^*)$. We use the subscripts ``B'' and ``S'' in reference to Bloch waves and to the hierarchical wave function proposed by Sutherland in \cite{sutherland} respectively, the reasons for that will be clear from what follows.
\par
The function $f_{\Lambda}$ from (\ref{df}) splits accordingly in two parts:
\begin{equation}
\label{fbfs}
f_{\Lambda}(\mathbf{x})=f_{\Lambda_{\rm B}}(\mathbf{x}) + f_{\Lambda_{\rm S}}(\mathbf{x}),
\end{equation}
where $\rmd f_{\Lambda_{\rm B}}$ and $\rmd f_{\Lambda_{\rm S}}$ are the pullbacks of 1-forms on $X_n$ from the classes $\Lambda_{\rm B}$ and $\Lambda_{\rm S}$ respectively. Moreover, since one can always choose a constant 1-form representative for $\Lambda_{\rm B}$, one can assume without loss of generality that $f_{\Lambda_{\rm B}}$ is a linear function:
\begin{equation}
\label{linform}
f_{\Lambda_{\rm B}}(\mathbf{x})=\frac{\mathbf{k}\mathbf{x}}{2\pi},
\end{equation}
for some $d\mbox{-dimensional}$ wave vector $\mathbf{k}$, which justifies the reference to Bloch states in its subscript.
\par
The second term in (\ref{fbfs}) grows much slower than $f_{\Lambda_{\rm B}}$, namely one can show that 
\begin{equation}
\label{loggrowth}
f_{\Lambda_{\rm S}}(\mathbf{x}) = \mathrm{O}\left(\log(|\mathbf{x}|)\right).
\end{equation}
The estimate (\ref{loggrowth}) can be easily understood in the case of self-similar structure models. The self-similarity is given by a dilatation  $\gamma_E: E \to E$ with a factor $a>1$, yielding a structure with an isomorphic hull. In other words, there exists an automorphism $\gamma: \Omega \to \Omega$ such that the following diagram commutes:
$$
\xymatrix{E \ar[r]^{\gamma_E} \ar[d]^\mu  &E \ar[d]^\mu\\
\Omega \ar[r]^\gamma & \Omega}
$$
The action of $\gamma$ carries to the first cohomology of $\Omega$:
$$
\gamma^*:\check{H}^1(\Omega) \to \check{H}^1(\Omega).
$$ 
The linear map $\gamma^*$ respects the splitting (\ref{split_cohom}), moreover, $\mathrm{coker}(\alpha_n^*)$ is an eigenspace of $\gamma^*$ with eigenvalue $1$ or $-1$. Since ${\gamma^*}^2(\Lambda_{\rm S})=\Lambda_{\rm S}$, the function 
$$f_{\Lambda_{\rm S}}(a^2\mathbf{x})- f_{\Lambda_{\rm S}}(\mathbf{x})$$ 
is a pullback of a continuous function on $X_n$ and is therefore bounded. Hence, $f_{\Lambda_{\rm S}}(\mathbf{x})$ grows at most as fast as $\log(|\mathbf{x}|)$. 
\par
Since the evolution of the wave function of the form (\ref{ansatz}) reduces to that of the pre-exponential factor $\Psi_0$ given by (\ref{psi0}), one might be tempted to conclude that all components of $\Lambda$ correspond to conserved quantities. However, it might happen that the same wave function admits the form (\ref{ansatz}) for more than one cohomology class $\Lambda$. This is indeed the case for the component $\Lambda_{\rm B}$, as follows from (\ref{linform}). Indeed, while $H^1(\T^N)$ is $N\mbox{-dimensional}$, $\Lambda_{\rm B}$ enters in (\ref{linform}) only through a $d\mbox{-dimensional}$ wave vector $\mathbf{k}$, and the ``perpendicular'' components of $\Lambda_{\rm B}$   give zero contribution to $f_{\Lambda}$. Furthermore, if the cocycle of the class $\Lambda$ has integer values on all 1-cycles of $X_n$, the exponential factor $\exp\left(2 \pi \rmi f_{\Lambda}(\mathbf{x})\right)$ is pattern equivariant and can thus be absorbed in the pre-exponential factor $\Psi_0$. In other words, the truly independent parameters of the ansatz are the factors
\begin{equation}
\label{floquet}
\lambda_i = \exp\left(2 \pi \rmi \Lambda(c_i)\right),
\end{equation}
where $c_i$ are the generators of the CW-homology group $H_1(X_n)$. The component $\Lambda_{\rm B}$ is thus not only reduced to the $d\mbox{-dimensional}$ wave vector $\mathbf{k}$, but is further factored over the equivalence relation
$$
\mathbf{k} \sim \mathbf{k}+\mathbf{k}_{\rm Bragg},
$$
where $\mathbf{k}_{\rm Bragg}$ are the wave vectors of the topological Bragg peaks \cite{kellendonk2013topological}. Since the latter form a dense module in $\R^d$, this makes $\Lambda_{\rm B}$ unusable as a classical continuous conserved quantity. 
\par
The situation is different for the component $\Lambda_{\rm S}$, for which the wave function (\ref{ansatz}) determines the factors (\ref{floquet}) unambiguously, as we shall see now. Recall that the space $X_n$ is constructed as a completion of a complement of $\T^N$ to $A_n$, an arrangement of thickened affine subtori of codimension 2. The CW-homology group $H_1(X_n)$ is generated by two types of cycles, by those inherited from $\T^N$ and by the loops around individual subtori of $A_n$. Each cycle of the second kind is an element of infinite order in $H_1(X_n)$, although this might not be true for their combinations, for $H_1(X_n)$ might contain torsion. Since on the other hand $\Lambda_{\rm B}$ vanishes on the cycles of the second kind, they are well suited to characterize the class $\Lambda_{\rm S}$. Let $c$ be a cycle of the second kind. Consider two points $x_1$ and $x_2$ on $\Omega$ such that $\pi(x_1)=\pi(x_2)$, but $\xi_n(x_1)$ and $\xi_n(x_2)$ are separated by the thickened subtorus of $A_n$ around which the cycle $c$ makes a loop. It is always possible to find a sequence of translations $d_m \in \R^d$ such that the following conditions are satisfied for $m>n$:
\begin{eqnarray*}
\xi_m(x_1+d_m)=\xi_m(x_2+d_m)\\
\xi_m(x_1-d_m)=\xi_m(x_2-d_m).
\end{eqnarray*}
Then the following map from $[-1,1]$ to $X_m$ defines a closed loop in $X_m$ having the same class as $c$:
\begin{equation}
\label{loop}
z \mapsto  \left\{
  \begin{array}{lr}
    \xi_m\left(x_1+(1+2z)d_m\right), & \mbox{if } -1 \le z <0\\
    \xi_m\left(x_2+(1-2z)d_m\right), & \mbox{if } 0 \le z \le 1
  \end{array}
\right.
\end{equation}
Let now $\psi_\Lambda(\mathbf(x), t)$ be a function of the form (\ref{ansatz}). 
Without loss of generality, at any given time $t$ the translations $d_m$ can be chosen in such a way that
\begin{eqnarray*}
\Psi_0(x_1 \pm d_m, t) \neq 0 \\
\Psi_0(x_2 \pm d_m, t) \neq 0. 
\end{eqnarray*}
Since $\mu(E)$ is dense in $\Omega$, there exist two sequences of points in the physical space $\mathbf{x}_{1,k} \in E$ and $\mathbf{x}_{2,k} \in E$, having as limit $x_1$ and $x_2$ respectively:
\begin{eqnarray*}
\lim_{k\to \infty} \mu(\mathbf{x}_{1,k})=x_1\\
\lim_{k\to \infty} \mu(\mathbf{x}_{2,k})=x_2
\end{eqnarray*}
(here both limits are considered in the topology of $\Omega$). In other words, the patterns surrounding $\mathbf{x}_{1,k}$ and $\mathbf{x}_{2,k}$ converge towards two different proximal singular tilings (see \cite{barge2013} for the definition of proximality). Note that $\Lambda(c)$ can be obtained as the integral of the form $\omega$ from (\ref{df}) over the loop (\ref{loop}). Since $\omega$ is continuous on $X_n$ this integral can be obtained as the limit of the contribution of two segments:
$$
\Lambda(c)=\lim_{k\to \infty} \left(
\int_{\xi_m\left(\mu(\mathbf{x}_{1,k} - d_m)\right)}
^{\xi_m\left(\mu(\mathbf{x}_{1,k} + d_m)\right)} \omega
+
\int_{\xi_m\left(\mu(\mathbf{x}_{2,k} + d_m)\right)}
^{\xi_m\left(\mu(\mathbf{x}_{2,k} - d_m)\right)} \omega
\right),
$$
where the integration is performed along the image of the physical space in $X_n$. The above integrals can be taken directly in $E$, yielding
$$
\Lambda(c)=
\lim_{k\to \infty} \left(
f(\mathbf{x}_{1,k} + d_m)-
f(\mathbf{x}_{1,k} - d_m)-
f(\mathbf{x}_{2,k} + d_m)+
f(\mathbf{x}_{2,k} - d_m)
\right).
$$
On the other hand, because of the continuity of $\Psi_0$ on $\Omega$ one has
$$
\lim_{m \to \infty}\frac{\Psi_0(x_2 \pm d_m, t)}{\Psi_0(x_1 \pm d_m, t)} = 1.
$$
Therefore
\begin{equation}
\label{lambda}
\exp\left(2 \pi \rmi \Lambda(c)\right) =
\lim_{m \to \infty}\left(
\lim_{k\to \infty} \left(\frac
{\psi_\Lambda(\mathbf{x}_{1,k} + d_m,t)}
{\psi_\Lambda(\mathbf{x}_{1,k} - d_m,t)}
\cdot
\frac
{\psi_\Lambda(\mathbf{x}_{2,k} - d_m,t)}
{\psi_\Lambda(\mathbf{x}_{2,k} + d_m,t)}
\right)\right).
\end{equation}
The above formula gives an explicit expression for the conserved factor (\ref{floquet}), defined for each cycle of the form (\ref{loop}). The value of the factor (\ref{lambda}) depends on the component $\Lambda_{\rm S}$ only, and the collection of these factors for all types of cycles $c$ of the second kind (that is, for all directions of ``worms'') gives the best possible characterization of  $\Lambda_{\rm S}$. 
\par
Recall now that the original motivation for the ansatz (\ref{ansatz}) was the quest of  generalized eigenstates of the stationary Schr\"odinger equation in quasicrystalline potentials. Would the generalized eigenstate have the form (\ref{ansatz}), what might be the values of the factor (\ref{floquet}) for it?  Because of the analogy with the Bloch states (\ref{bloch2}), one might be tempted to constrain the factors (\ref{floquet}) to unitary complex numbers, leaving non-unitary values to the description of evanescent waves. Note however, that because of the logarithmic growth rate of $f_{\Lambda_{\rm S}}$ (\ref{loggrowth}), the growth of the wave function (\ref{ansatz}) will be bounded by a power law even if the factor (\ref{floquet}) is not unitary. By virtue of Schnol theorem \cite{cycon1987}, this makes such functions perfectly acceptable as generalized eigenstates. This fact was first observed by Sutherland in \cite{sutherland}, who proposed a hierarchical wave function with purely real factor (\ref{floquet}) as a ground state for a modified tight binding model on Penrose tiling. As we shall see below, this is also the case for the ground state for a variety of other tight-binding models.
\section{Tight-binding models}\label{sec:tight-binding}
The band theory of crystalline solids may be formulated equally in continuous space or in tight binding models. The latter are traditionally derived from the former by first projecting the full Hamiltonian to a single spectral band and then decomposing it in the basis of Wannier functions \cite{ashcroft1976solid}. However, since the tight binding models capture the essential features of the crystalline band structure, they are often considered independently of the underlying continuous space models. By extension, electrons in quasicrystalline potentials are also modelled by tight binding Hamiltonians, even though there is no known way to generalize the Wannier construction for this case. Such models are built upon tilings of the physical space $E$, where the ``atoms'' are usually represented by tile vertices and the valence bonds correspond to the edges of the tiling.
\par
Sutherland discovered an example of a hierarchical state for a tight-binding model on Penrose tiling \cite{sutherland} (his model was obtained from the standard one by adding an on-site energy term tailored to make the proposed wave function an exact eigenstate). The ansatz of  \cite{sutherland} was based on observation that the de Bruijn arrows \cite{de1981algebraic, de1988symmetry} make up a curl-free (co)vector field defined on the edges of the tiling. One can reformulate this in terms of the cellular decomposition of $E$ defined by the tiling. Let $T_k \subset E$ stands for the union of closed $k\mbox{-dimensional}$ faces (or the $k\mbox{-skeleton}$) of the tiling $T$. Then, the de Bruijn arrows define a strongly pattern equivariant 1-cocycle on $T_1$. The curl-free property signifies that this cocycle is a coboundary in ordinary cellular complex of $T_1$ (although it is not in the pattern equivariant cellular complex). There exists a systematic way to construct such cocycles from \v{C}ech cohomology classes of the tiling hull $\Omega$. In fact, according to \cite{sadun2007pattern}, the integer PE cohomologies of $T$ are isomorphic to \v{C}ech cohomologies of $\Omega$. We shall use this isomorphism below to formulate the discrete version of the ansatz (\ref{ansatz}). 
\par
Let us now extend the results of Section \ref{sec:ansatz} for the case of discrete models. We shall study the time-dependent tight-binding Schr\"odinger equation:
\begin{equation}
\label{schrodinger-tb}
\rmi \dot \psi_\Lambda(p, t)= \sum_q H_{pq} \psi_\Lambda(q, t),
\end{equation}
where $p, q \in T_0$ are the vertices of the tiling. We also assume that the Hamiltonian is of finite range, so that the sum in (\ref{schrodinger-tb}) is finite. The matrix elements $H_{pq}$ are strongly pattern equivariant, more specifically, there exists a positive integer $n$ such that whenever the patches of tilings around the pairs of vertices $(p_1, q_1)$ and $(p_2, q_2)$ agree up to the $n\mbox{-th}$ corona \cite{sadun2003tiling}, one has $H_{p_1 q_1}= H_{p_2 q_2}$. 
\par
Let $\gamma$ be a PE 1-cocycle of the tiling corresponding the cohomology class $\Lambda \in \check{H}^1(\Omega)$. The tight-binding wave function $\psi_\Lambda(p)$ is a direct generalization of (\ref{ansatz}):
\begin{equation}
\label{ansatz-tb}
\psi_\Lambda(p,t)=\Psi_0\left(\mu(p), t \right) \exp\left(2 \pi \rmi f_{\Lambda}(p)\right),
\end{equation}
where now $\rmd f_{\Lambda}=\gamma$ in the sense of ordinary cellular complex of the tiling (note that $f_\Lambda$ {\em is not} pattern equivariant if $\Lambda \neq 0$). So far we assume that the function $\Psi_0(y, t)$ is defined for $y \in \mu(T_0)$ only, however, our goal is to continue it to the closure of $\mu(T_0)$ in the hull topology:
$$
\Xi=\overline{\mu(T_0)}
$$
(the space $\Xi$ is called canonical tiling transversal in \cite{kellendonk2003pattern,bellissard2006spaces}). For the moment, we shall not make any assumption about regularity of $\Psi_0$ and shall consider its evolution on $\mu(T_0)$ only:
\begin{equation}
\label{cauchy_tb}
\rmi\dot\Psi_0(\mu(p),t)=\sum_q\tilde H_{pq} \Psi_0(\mu(q),t),
\end{equation}
where $\tilde H_{pq}$ is a ``gauge transformed'' Hamiltonian:
\begin{equation}
\label{hpq}
\tilde H_{pq}=H_{pq} \exp\left(2 \pi \rmi \left(f_{\Lambda}(q)-f_{\Lambda}(p)\right)\right).
\end{equation}
Note that since $f_{\Lambda}$ need not to be real (see discussion at the end of Section \ref{sec:ansatz}), $\tilde H_{pq}$ may be not hermitian.
\par
So far the index $n$ used to enumerate the spaces $X_n$ in the sequence (\ref{lim_hull}) had no particular significance; the only thing that counted was the existence of the inverse limit. However, when working with discrete models, there is a somewhat natural way to choose the spaces $X_n$. To make this point clear it is convenient to introduce a coarser topology on the tiling transversal. Let $D_n$ be the closure of $\xi_n(\Xi)$ in the topology of $X_n$:
\begin{equation}
\label{dn}
D_n=\overline{\xi_n(\Xi)}.
\end{equation} 
The spaces $D_n$ are finite disjoint unions of $K_n$ polytopes $D_{n,k}$ of dimension $N-d$:
\begin{equation}
\label{dnk}
D_n=\bigcup_{k=1}^{K_n} D_{n,k},
\end{equation}
where each polytope is an acceptance domain of a finite patch of the tiling. As $n$ grows, the corresponding patches become bigger and the connected components $D_{n,k}$ become smaller, and finally one recovers the Cantor set topology of $\Xi$ as an inverse limit of $D_n$. Furthermore, the size of $D_{n,k}$ decreases uniformly. Indeed, by virtue of Liouville theorem on Diophantine approximations, the quadratic irrationality conditions imply that any linear dimension of $D_{n,k}$ is bounded from below by $C n^{-1}$ for some $C>0$. Therefore there exist constants $C_1>C_2>0$ such that:
\begin{equation}
\label{uniform}
C_1 >\mu_{\rm T}(D_{n,k})n^{N-d}>C_2,
\end{equation}
where $\mu_{\rm T}$ is a translation invariant transverse measure on $\Xi$ (one can carry $\mu_{\rm T}$ to $D_n$ since the natural map $\Xi \to D_n$ is one-to-one almost everywhere). Let us call the sequence of spaces $X_n$ {\em compatible} with the Hamiltonian if for any locally constant function $\chi_1$ on $D_n$ there exists a locally constant function $\chi_2$ on $D_{n+1}$ such that
\begin{equation}
\label{compatible}
\sum_q H_{pq} \chi_1\left(\mu(q)\right)=\chi_2\left(\mu(p)\right).
\end{equation}
For a given $X_1$ we call a sequence $X_n$ {\em tight} if it is compatible with the Hamiltonian and for $n>2$ the spaces $D_n$ have the coarsest possible topology. Notice that if the sequence $X_n$ is compatible with the Hamiltonian and tight, it agrees with the metric of ``approximate match'' $D(\mathbf{x}, \mathbf{y})$ introduced in Section \ref{sec:qc_hulls} in the sense that the distance between the vertices belonging to the same connected component of $D_n$ scales as $1/n$. More precisely, for any two vertices $p,\,q \in T_0$ such that $\xi_n(\mu(p))$ and $\xi_n(\mu(q))$ belong to the same connected component of $D_n$ but $\xi_{n+1}(\mu(p))$ and $\xi_{n+1}(\mu(q))$ belong to different connected components of $D_{n+1}$, one has $D(p, q) \sim 1/n$. In what follows we shall always assume that $X_n$ is compatible with the Hamiltonian and tight. 
\par
\begin{figure}[h]
  \centering
    \includegraphics[width=0.8\textwidth]{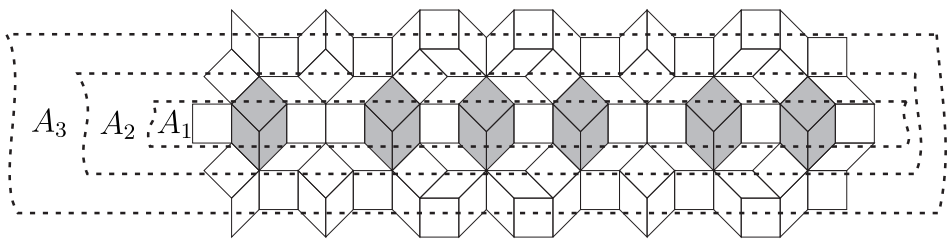}
    \caption{A patch of Ammann-Beenker tiling in a singular position containing an infinite ``worm''. Simultaneous flipping upside down of all shaded hexagons produces another singular tiling. The bands bounded by dashed lines represent the footprints $(\pi \circ \mu)^{-1}(A_n)$ of the ``forbidden sets'' $A_n$ on the physical space $E$. The corresponding sequence of spaces $X_n$ is compatible with the Hamiltonian since all nearest neighbours of the vertices from $(\pi \circ \mu)^{-1}(A_n)$ belong to $(\pi \circ \mu)^{-1}(A_{n+1})$. The sequence $X_n$ is also tight since the bands cannot be made narrower without loss of compatibility with the Hamiltonian.}
    \label{fig:worm}
\end{figure}
\begin{figure}[h]
        \begin{center}
        \subfigure[]{\label{fig:d1}\includegraphics[width=0.45\textwidth]{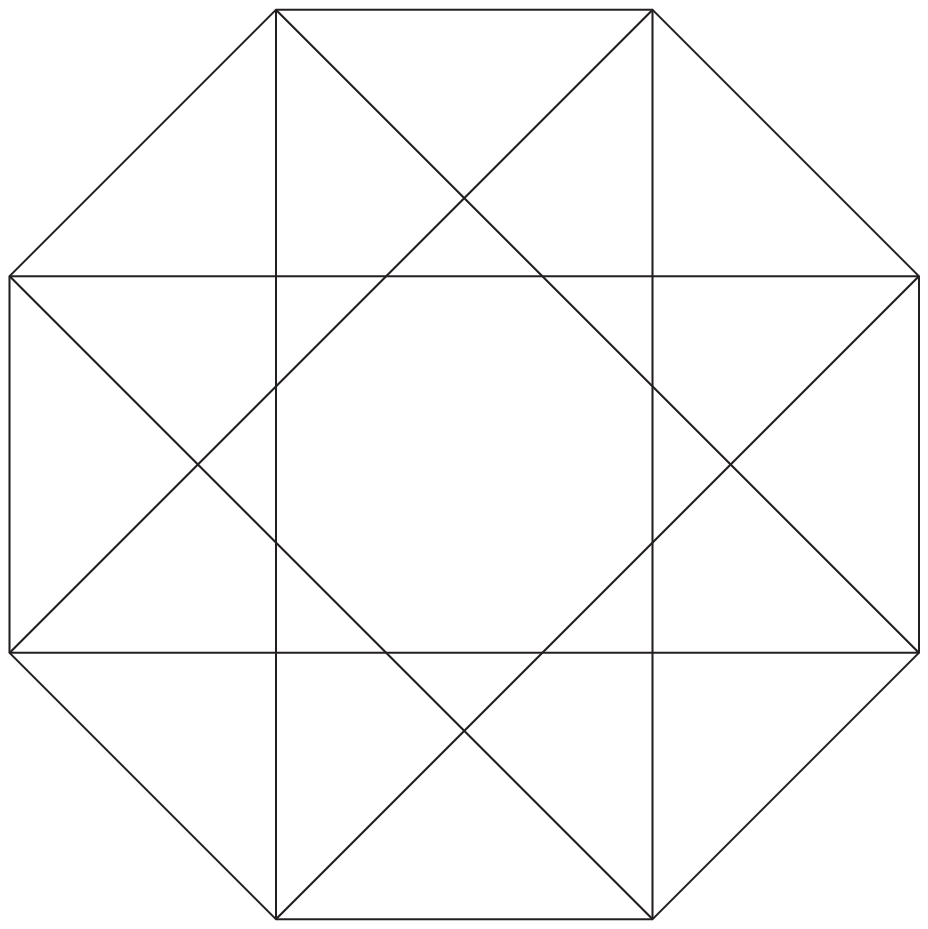}}
        \subfigure[]{\label{fig:d2}\includegraphics[width=0.45\textwidth]{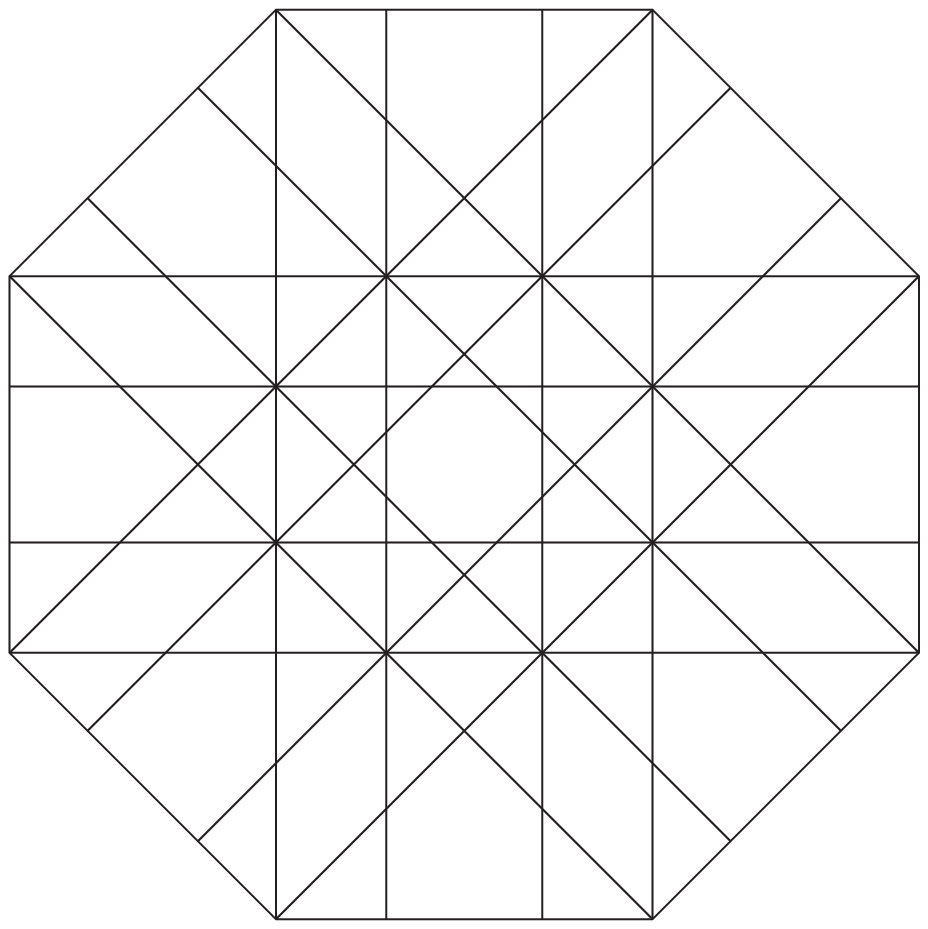}}
        \end{center}
        \caption{The spaces $D_n$ (\ref{dn}) of the Ammann-Beenker tiling for $n=1$ \subref{fig:d1} and $n=2$ \subref{fig:d2}. Topologically, $D_n$ are disjoint unions of closed polytopes.}\label{fig:c0} 
\end{figure}

Let us illustrate the above by the example of Amman-Beenker tiling. Consider the situation when $\mu$ maps $E$ to the singular set of $\Omega$ (we shall say that the corresponding tiling is in a singular position). In this case, the cut $\pi(\mu(E))$ intersects the ``thickened'' arrangement $A_n$. The corresponding inverse image $(\pi \circ \mu)^{-1}(A_n)$ represents a footprint of $A_n$ on the physical space $E$. In the general case this set is a straight band aligned along an infinite ``worm'', although it might also consist of several intersecting bands. For the sake of simplicity let us assume that there is only one infinite ``worm'' in the tiling, and therefore $(\pi \circ \mu)^{-1}(A_n)$ consists of a single band. This situation is illustrated on Figure \ref{fig:worm}. Simultaneous flipping of all shaded hexagons on Figure \ref{fig:worm} produces another singular tiling. The vertices of these two tilings correspond to distinct points in $\Omega$, but they are partially glued together in the topology of $X_n$. More specifically, the map $\xi_n \circ \mu$ distinguishes the points of both tilings (the flipped and the original one) if and only if they belong to the band $(\pi \circ \mu)^{-1}(A_n)$. The vertices of the tilings lying within $(\pi \circ \mu)^{-1}(A_n)$ correspond to the boundaries of the connected components of $D_n$ (\ref{dnk}), as illustrated by Figure \ref{fig:c0}. The sequence of spaces $X_n$ is compatible with the Hamiltonian if and only if the band $(\pi \circ \mu)^{-1}(A_{n+1})$ is large enough to contain all nearest neighbours of the vertices from $(\pi \circ \mu)^{-1}(A_n)$. The dashed bands on Figure \ref{fig:worm} satisfy this conditions. Moreover, the corresponding sequence $X_n$ is also tight, since the dashed bands are just as large as needed to make it compatible with the Hamiltonian.
\par
We are now ready to impose some regularity conditions on $\Psi_0$. Let us consider the Hilbert space $L^2(\Xi, \mu_{\rm T})$. Let $\mathcal{W}_n$ stand for the finite dimensional subspace of $L^2(\Xi, \mu_{\rm T})$ spanned by pullbacks of locally constant functions on $D_n$ by $\xi_n$. Note that $\mathcal{W}_n$ form a  growing sequence of Hilbert spaces:
\begin{equation}
\label{embedding}
\mathcal{W}_n \subset \mathcal{W}_{n+1}.
\end{equation}
Let $\mathcal{V}_n$ stand for the orthogonal complement to $\mathcal{W}_{n-1}$ in $\mathcal{W}_n$ (we set $\mathcal{V}_1=\mathcal{W}_1$). Therefore, $\mathcal{W}_n$ decomposes in a finite Hilbert sum:
\begin{equation}
\label{sum_of_vs}
\mathcal{W}_n =\bigoplus_{k=1}^n \mathcal{V}_k.
\end{equation}
We shall now construct a weighted Hilbert space of functions on $\Xi$. Given  a sequence of weights $r_i>0$, one can introduce a weighted Hilbert norm $\| \cdot \|_r$ on $\mathcal{W}_n$ in the following way. Because of (\ref{sum_of_vs}), any vector $w \in \mathcal{W}_n$ decomposes as 
$$
w=\sum_{i=1}^n v_i,
$$
where $v_i \in \mathcal{V}_i$. We define the $\| \cdot \|_r\mbox{-norm}$ of $w$ as
\begin{equation}
\label{r-norm}
\| w \|_r^2 = \sum_{i=1}^n r_i \| v_i \|^2,
\end{equation}
where $\| \cdot \|$ is the norm on $\mathcal{V}_i$ inherited from $L^2(\Xi, \mu_{\rm T})$. The norms (\ref{r-norm}) agree on the embeddings (\ref{embedding}) and provide the union of all $\mathcal{W}_n$ with the structure of a pre-Hilbert space. We shall denote the corresponding Hilbert space by $\mathcal{H}(r)$. 
\par 
The construction of the weighted Hilbert space $\mathcal{H}(r)$ bears obvious resemblance to that of Sobolev spaces. Indeed, for growing sequences of weights $r_n$ the functions from $\mathcal{H}(r)$ behave more regularly than just square integrable ones. Moreover, if the growth is fast enough, they are continuous on $\Xi$, as can be seen from the following arguments. According to (\ref{uniform}), the $\mu_{\rm T}\mbox{-measure}$  of smallest contiguous pieces of $D_n$ scales with $n$ as $n^{d-N}$ and therefore the $L^\infty$ norm of the vectors from the unit ball in $\mathcal{H}(r)$ belonging to $\mathcal{V}_n$ is bounded from above by $C r_n^{-1/2} n ^{(N-d)/2}$ for some $C>0$. Hence, if the following series converges:
\begin{equation}
\label{r_convergence}
\sum_{n=1}^\infty r_n^{-1} n^{N-d} < \infty,
\end{equation}
the functions from $\mathcal{H}(r)$ are continuous on $\Xi$. If the growth of the weights follows a power law this is the case when
\begin{equation}
\label{r_power_law}
r_n^{-1} =\mathrm{O}(n^{-\alpha})\quad\mbox{where }\alpha> N-d+1.
\end{equation}
\par
So far the topology of $\Xi$ did not play any role in the definition of $\Psi_0$ in (\ref{cauchy_tb}). Let us now consider (\ref{cauchy_tb}) as a Cauchy problem with the initial condition $\Psi_0(y, 0)$ set by a function from $\mathcal{H}(r)$, where the latter is defined with a sequence of weights $r_n$ satisfying (\ref{r_convergence}). This condition is essential since it allows one to consider the elements of $\mathcal{H}(r)$ as {\em bona fide} functions on $\Xi$, having a well defined value at each point. This would not be possible had we used $L^2(\Xi, \mu_{\rm T})$ instead of $\mathcal{H}(r)$. In fact, an element of $L^2(\Xi, \mu_{\rm T})$ corresponds to a {\em class} of functions on $\Xi$ and speaking of its values on a zero measure subset such as $\mu(T_0) \subset \Xi$ is meaningless. However, since (\ref{cauchy_tb}) defines the time evolution of $\Psi_0(y, t)$ for $y \in \mu(T_0)$ only, a question naturally arises whether $\Psi_0(y, t)$ can be extended to a continuous function on $\Xi$ at a time $t>0$. Let us show that this is indeed the case if the weights $r_n$ satisfy the following inequality:
\begin{equation}
\label{bound_on_r}
A^{-1}<r_{n+1}/r_{n} < A
\end{equation}
for some real constant $A>0$. Consider a case when $\Psi_0 \in \mathcal{W}_n$. Since the sequence $X_n$ is supposed compatible with the Hamiltonian, the right-hand side of (\ref{cauchy_tb}) is a pullback of a vector from $\mathcal{W}_{n+1}$. Therefore, the right-hand side of (\ref{cauchy_tb}) defines an operator $\mathcal{W}_n \to \mathcal{W}_{n+1}$, which is uniformly bounded for all $n$ because of 
(\ref{bound_on_r}). The limit of these operators in the strong operator topology yields a bounded operator on $\mathcal{H}(r)$, which we shall denote $\bar H$. The equation (\ref{cauchy_tb}) can therefore be rewritten as an evolution of a vector of $\mathcal{H}(r)$:
\begin{equation}
\label{evol_hilbert}
\rmi \dot \Psi_0 = \bar H \Psi_0.
\end{equation}
Although the operator $\bar H$ is not hermitian on $\mathcal{H}(r)$, the holomorphic functional calculus yields a bounded evolution operator $\exp(-\rmi \bar H t)$. Therefore, if the initial condition for $\Psi_0$ in (\ref{evol_hilbert}) is set by a function from $\mathcal{H}(r)$, the solution will remain in $\mathcal{H}(r)$ forever.
\section{Numerical results}\label{sec:numerical}
The wave function (\ref{ansatz}) bears a manifest resemblance to Bloch states (\ref{bloch}), with the conserved parameter $\Lambda$ playing the role of the quasimomentum $\mathbf{k}$. However, in the case of crystals, one has something more than just a conservation of the quasimomentum. Indeed, the Schr\"odinger operator with periodic potentials admits a complete set of generalized eigenstates having the form (\ref{bloch}). A question naturally arises whether in the case of quasicrystals the ansatz (\ref{ansatz}) could provide such a set. In this section we report numerical results in favour of this hypothesis.
\par 
A traditional way of studying numerically eigenstates in tight-binding models consists of considering a finite patch of the structure with some sort of boundary conditions. The Hamiltonian then becomes a finite matrix, and the corresponding eigenvectors are interpreted as an approximation to the generalized eigenstates of the infinite model. However, this approximation might be rather poor, in contrast to that of the spectral characteristics of the model, such as the integrated density of states. Indeed, in the general case, the restriction of any of the generalized eigenstates of the infinite model to the patch does not produce an eigenvector of the corresponding finite matrix, even if the energy of the generalized eigenstate coincides with the one of the matrix eigenvalues. Typically one obtains a linear combination of many eigenvectors, with significant contribution coming from about $g$ of them, where $g$ is the so-called ``Thouless number'' \cite{abrahams1979scaling} (or the dimensionless conductance) of the patch. The notorious exception is the case of periodic lattices models with periodic boundary conditions, where the eigenvectors are in fact Bloch states restricted to the patch considered. This makes clear the role played by the conditions at the boundary of the patch in this approach, especially in the case where the dimension of the physical space is larger than one (since in the one-dimensional case one always has $g<1$). 
\par
As follows from the above, a successful numerical approximation of a generalized eigenstate in a quasiperiodic tiling is possible only if the boundary conditions respect the structure of the eigenstate --- in the same way the periodic boundary conditions respect the structure of Bloch states in crystals. Clearly, we should use the compatibility with the suggested form of the eigenstate (\ref{ansatz}) as a criterion for the choice of the boundary conditions. Unfortunately, unlike for the Bloch states, no boundary conditions are compatible with all functions of the form (\ref{ansatz}) simultaneously. It is still possible, however, to find boundary conditions compatible with one chosen generalized eigenstate. In this situation, the ground state is the natural choice. First of all, for time-reversal Hamiltonians this state is easily identifiable, as no other generalized eigenstate can be made real and non-negative everywhere. Moreover, in the case of tilings with rotational symmetry, this state can be chosen symmetric, which would correspond to $\Lambda_{\rm B}=0$, leaving unknown only $\Lambda_{\rm S}$.
\par
We have chosen two models for the numerical study, the octagonal Ammann-Beenker tiling and the Penrose tiling. The tight-binding models on these two tilings became {\em de facto} standard benchmarks for the problem considered. Unlike Sutherland \cite{sutherland}, we use these models with the standard parameters, that is with zero on-site energy and hopping integrals all equal to 1. Apart of their popularity, the other advantage of Ammann-Beenker and Penrose tilings is that for both of them the group $\check{H}^1(\Omega)$ has rank 5. Since 4 of these dimensions come from $H^1(\T^N)$ in (\ref{split_cohom}), this leaves only one dimension for the class $\Lambda_{\rm S}$. Therefore, if the ground state of these models have the form (\ref{ansatz}), it can be completely defined by a function $\Psi_0 \in C(\Xi)$ and a single phase factor (\ref{lambda}) (which should probably be called a scaling factor since, as we shall see, $\lambda$ is a real number for the case considered).
\par
\begin{figure}[h]
  \centering
    \includegraphics[width=0.8\textwidth]{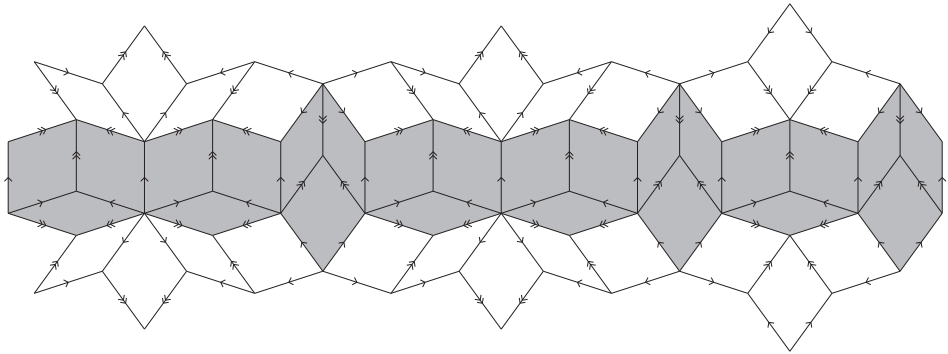}
    \caption{A patch of decorated Penrose tiling in a singular position. The gray area represents a ``worm''. Flipping the ``worm'' upside down changes the count of arrows along any path crossing the ``worm''. Note that only the count of single arrows is affected.}
    \label{fig:penrose_worm}
\end{figure}

The construction of the ansatz (\ref{ansatz-tb}) requires a pattern-equivariant 1-cocycle on the 1-skeleton of the tiling. Luckily, the arrows of the standard decorations of both Ammann-Beenker and Penrose tilings provide us with exactly what we need (the fact that the de Bruijn decorations \cite{de1981algebraic, de1988symmetry} represent a cocycle was first discovered by Sutherland \cite{sutherland}, who remarked that both single and double arrows form irrotational vector fields on the tiling). It remains to verify that this cocycle is not trivial and that it does not contain components from $H^1(\T^N)$ in the decomposition (\ref{split_cohom}). To verify the latter is suffices to consider the action of the point symmetry group on the tiling by rotations. Since both decorations are invariant with respect to rotations, the component $\Lambda_{\rm B}$ of the corresponding cocycle must be zero. The non-triviality of the cocycle can be verified by considering how the count of arrows along an open path crossing a ``worm'' is affected by flipping the latter. Figure \ref{fig:penrose_worm} illustrates this idea. One can note that flipping the ``worm'' upside down changes the count of simple arrows by $\pm 2$, and does not modify the count of double arrows. This comes as no surprise, for the group $\mathrm{coker}(\alpha_n^*)$ in (\ref{split}) has rank 1 for the Penrose tiling, and therefore one of the types of the arrows is redundant. We shall thus use the count of simple arrows to construct the function $f_{\Lambda}(p)$ in (\ref{ansatz-tb}) (note that the decoration of the Penrose tiling used in \cite{sutherland} differs from that proposed by de Bruijn, namely the directions of the arrows are reversed and the single and the double arrows permuted).
\subsection{The boundary conditions}\label{sec:boundary}
The hulls of both Ammann-Beenker and Penrose tilings have symmetry of dihedral groups $\mathbf{D}_8$ and $\mathbf{D}_{5}$ respectively (actually, the symmetry of the hull of the undecorated Penrose tiling is $\mathbf{D}_{10}$, but the arrows break it down to $\mathbf{D}_5$). If the ground state has the form (\ref{ansatz-tb}), the pre-exponential factor $\Psi_0$ and the cohomology class $\Lambda$ should both have the symmetry of the hull. Thus, it would be natural to choose boundary conditions that impose the symmetry of the hull on the wave function. Note, however, that the tiling itself does not necessarily exhibit this symmetry, and even if it does, the corresponding fundamental domain is unbounded. One can however take into account the repetitivity of the tiling and use an approximate local symmetry instead of the perfect global one. We follow this approach and choose the patches of triangular form bounded by local mirror lines. More precisely, the local mirror symmetry means that for a patch of linear size $L$ the tiling possesses a mirror symmetry in the vicinity of each edge of the triangle up to a distance $cL$ from the edge, where $c>0$ is some constant. Triangular patches possessing these properties can be conveniently generated by successive inflations. 
\par
In the continuous case the local mirror symmetry can be imposed by Neumann boundary conditions (zero normal component of $\nabla \psi$). A similar condition for tight-binding models is slightly more involved. One can start with Ammann-Beenker tiling and a patch of the form shown on Figure \ref{fig:triangle_a}. In this case, the triangular patch is in fact a fundamental domain of the action of a global symmetry group $p4m$ on some periodic tiling. Then the symmetric part of the tight-binding Hamiltonian for this tiling yields an hermitian operator acting on a finite-dimensional Hilbert space spanned by the vertices of the patch of Figure \ref{fig:triangle_a}. One can easily check that this operator is also tight-binding in the sense that its non-zero matrix elements correspond to the edges of the patch of the tiling. The matrix elements of inner edges are the same as for the original Hamiltonian, but that of the edges having at least one end at one of the bounding mirror lines are modified. This modification can be naturally interpreted as a tiling version of Neumann boundary conditions. Unfortunately, this approach does not work for Penrose tiling, for which none of the possible triangular patches is a fundamental domain of a plane crystallographic group.
\par
A careful examination of the modified matrix elements in the previous case shows that the alteration depends on the local environment of the corresponding link only. Therefore, instead of considering the problem in a triangular patch (like the one of Figure \ref{fig:triangle_p}) one can study it first in a semi-infinite open angle of measure $2 \pi x$ (for some rational $x$), bounded by two mirror lines. In the case of Penrose tiling, this angle consists of one or several fundamental domains of the group $\mathbf{D}_5$, and one may try to apply the same reasoning as above. The results can be conveniently formulated in the following way. Let $s(p)$ stand for the part of the plane in the vicinity of the vertex $p$ belonging to the interior of the angle. That is $s(p)=1$ if $p$ is an interior point of the angle, $s(p)=1/2$ if $p$ lies at the edge and if $p$ is a vertex of the  angle, $s(p)=x$. Let also $s(p,q)$ stand for the similar quantity for inner points of the interval $(p,q)$ (in two dimensions $s(p,q)$ always equals either $1$ or $1/2$, but the construction can be generalized to higher dimensions where this is not always the case). Then, if no edge of the tiling crosses the mirror line, the mirror boundary conditions correspond to the following modification in the matrix elements of the Hamiltonian:
\begin{equation}
\label{neumann}
H'_{pq}=\frac{s(p,q)}{\sqrt{s(p)s(q)}} H_{pq},
\end{equation}
where $H'$ acts on the states inside the angle only. One can show that if $\psi(p)$ is an eigenstate of $H$ having the symmetry of the finite reflection group generated by the sides of the angle, then 
\begin{equation}
\label{weight_psi}
\psi'(p)=\sqrt{s(p)} \psi(p)
\end{equation}
is an eigenstate of $H'$ with the same eigenvalue. 
\par
\begin{figure}[!h]
  \centering
    \includegraphics[width=0.8\textwidth]{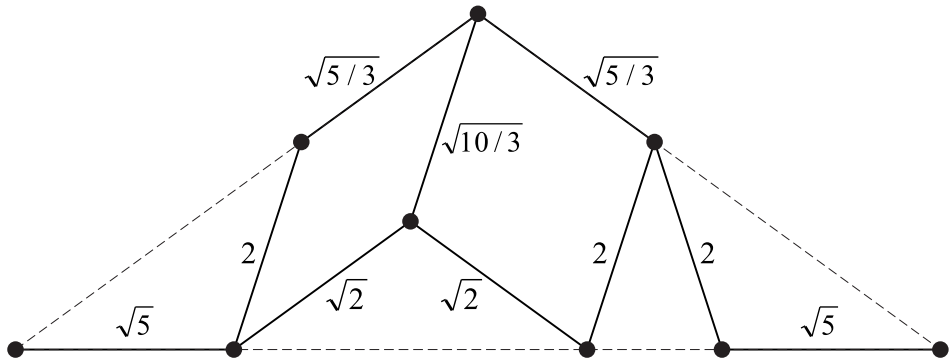}
    \caption{The values of the matrix elements of the tight-binding Hamiltonian on Penrose tiling modified by the weight factors (\ref{neumann}) to reflect the mirror boundary conditions at the edges of a triangular patch of the tiling. Since this patch has no inner edges, all matrix elements are altered.}
    \label{fig:penrose_weights}
\end{figure}

One can remark that the modification of the matrix elements in (\ref{neumann}) depends only on the local environment of the edge $(p,q)$. Therefore one can use (\ref{neumann}) as an expression of the mirror boundary conditions in finite triangular patches of the tiling as well. An example of such modification is shown on Figure \ref{fig:penrose_weights}.
\subsection{The ground state}
In order to test numerically whether the ground-state has the form of the ansatz (\ref{ansatz-tb}) with $\Psi_0$ continuous on $\Xi$, one has to know the exponential factor $\exp(2 \pi \rmi f_\Lambda(p))$. The latter can be readily obtained from the value of $\lambda$ in (\ref{floquet}), which can be approximately determined by the ground state $\psi_\Lambda$ itself according to (\ref{lambda}). However, since we expect that the ground state has the symmetry of the hull, and therefore has $\Lambda_{\rm B}=0$, it is possible to use only two points to estimate $\lambda$ instead of four points used in (\ref{lambda}). Indeed, one can choose two vertices $a$ and $b$ of the tiling in such a way that the the distance between $\mu(a)$ and $\mu(b)$ is small (that is, the tilings around $a$ and $b$ agree up to a large distance), and that the net count of arrows (the simple ones in the case of the Penrose tiling) between $a$ and $b$ equals 2. The first condition means that $\xi_n \circ \mu$ maps the path connecting $a$ and $b$ to an almost closed loop in $X_n$, and the second one signifies that this loop makes exactly one turn around one of the subtori of $A_n$, and can therefore be considered as an approximation to the cycle $c_k$ in (\ref{floquet}). Therefore, the factor $\lambda$ can be estimated as
\begin{equation}
\label{estim_lambda}
\lambda \approx \frac{\psi(a)}{\psi(b)}.
\end{equation}
This formula can be used to estimate the value of $\lambda$ numerically for the ground state wave function in a finite patch. Note however that if $a$ and $b$ are not inner points of the patch one has to modify (\ref{estim_lambda}) by taking into account the weight factor of (\ref{weight_psi}). 
\begin{figure}[!h]
  \centering
    \includegraphics[width=\textwidth]{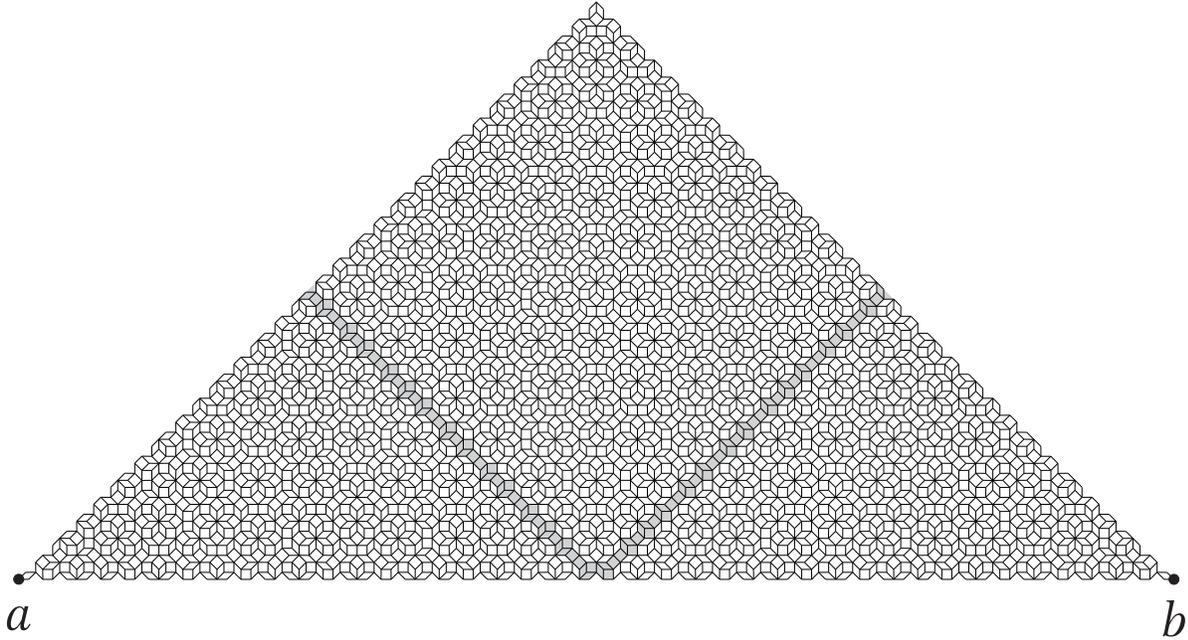}
    \caption{A triangular patch of Ammann-Beenker tiling with 4180 sites bounded by local mirror symmetry lines. The local environments at the points $a$ and $b$ agree up to the distance to the nearest shaded ``worm''. The net count of Ammann arrows between $a$ and $b$ equals 2.}
    \label{fig:triangle_a}
\end{figure}
\begin{figure}[!h]
  \centering
    \includegraphics[width=\textwidth]{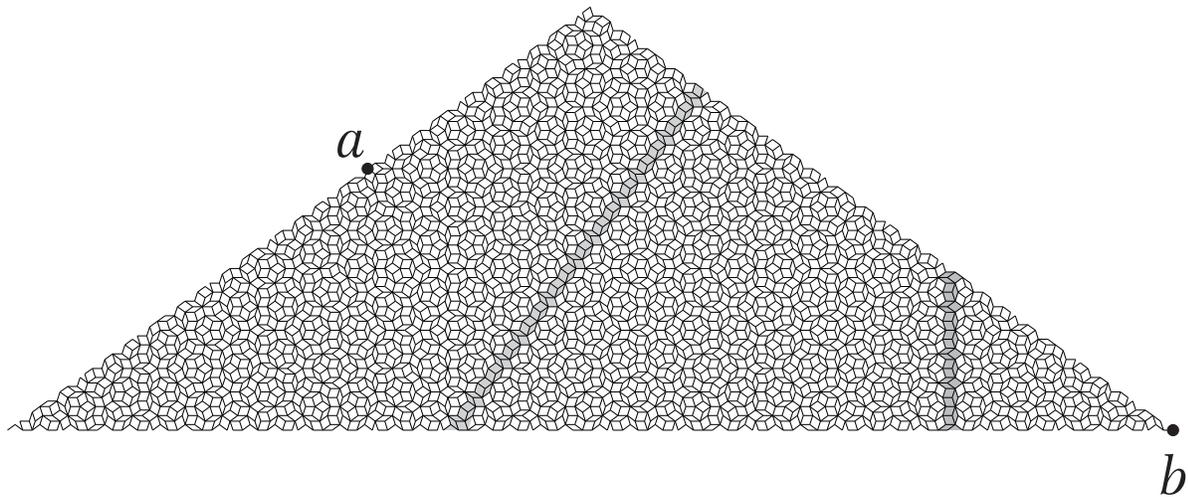}
    \caption{A triangular patch of Penrose tiling with 3500 sites bounded by local mirror symmetry lines. The local environments at the points $a$ and $b$ agree up to the distance to the nearest shaded ``worm''. The net count of simple de Bruijn arrows between $a$ and $b$ equals 2. Note that since $s(a) \neq s(b)$ one has to apply the weight factors to the estimate (\ref{estim_lambda}): $\lambda \approx 5^{-1/2}\psi(a)/\psi(b)$ }
    \label{fig:triangle_p}
\end{figure}
\par
The figures \ref{fig:triangle_a} and \ref{fig:triangle_p} illustrate the above construction. Note that the relative positions of $a$ and $b$ within the triangular patch are the same for all sizes of the triangle. Indeed, since the consecutive inflations bring $\mu(a)$ and $\mu(b)$ closer together in the hull, the tiling around $a$ and $b$ agree up to a distance comparable to the size of the patch. Moreover, this choice allows to hope a kind of regularity in the behaviour of the consecutive approximation in view of possible extrapolation of the results. This is indeed the case, since, as can be seen from the tables \ref{tab:ammann} and \ref{tab:penrose}, the differences between the values of $E_0$ and $\lambda$ for every second stage of inflation fall off in approximately geometric progression. The parameters of the ground state for the infinite tiling can be then estimated by means of the Richardson extrapolation \cite{richardson1927deferred},  yielding the
results shown in the bottom row of the tables \ref{tab:ammann} and \ref{tab:penrose}. The uncertainty of the estimate is evaluated by comparing the convergence for different shapes of the patch.
\begin{table}
\caption{\label{tab:ammann}The ground state energy $E_0$ and the corresponding scaling factor $\lambda$ (\ref{floquet}) estimated for the finite patches of Ammann-Beenker tiling with local mirror boundary conditions for different number of sites. The figures at the bottom row are obtained by Richardson extrapolation.} 
\begin{indented}
\lineup
\item[]\begin{tabular}{@{}*{3}{l}}
\br                              
Number of sites & $-E_0$ & $\lambda$\\
\mr
30 & 4.22131345474597 & 1.31023580279858 \\
141 & 4.22169077249007 & 1.36478815968518 \\
747 & 4.22169711324415 & 1.35739594032671\\ 
4180 & 4.22169745156009 & 1.35821790197493\\
23950 & 4.22169745684341 & 1.35805783747795\\
138601 & 4.22169745712397 & 1.35808037029283\\
$\infty$ & 4.2216974571286(2) & 1.358076(2)\\
\br
\end{tabular}
\end{indented}
\end{table}
\par
\begin{figure}[!h]
  \centering
    \includegraphics[width=0.8\textwidth]{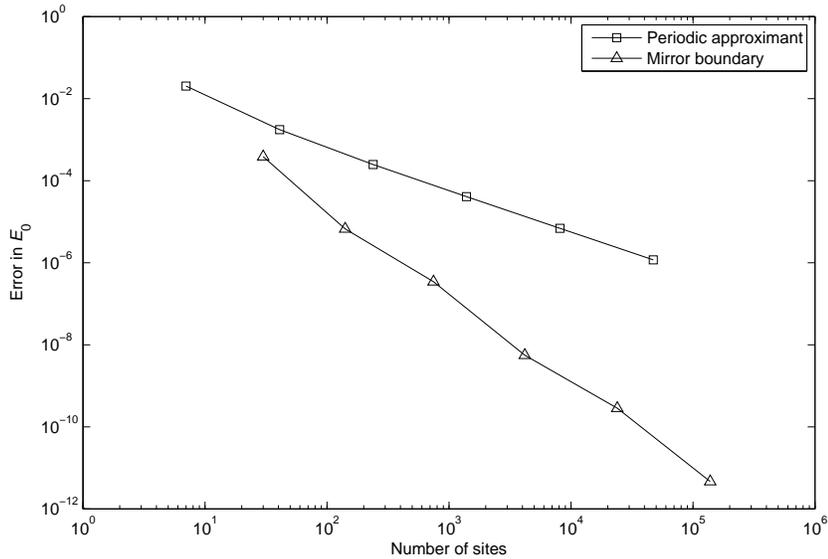}
    \caption{The error in estimation of the ground state energy as a function of the number of sites in the finite patch of the Ammann-Beenker tiling. The convergence rate with the local mirror boundary conditions is much higher that with the periodic boundary condition for standard approximants \cite{duneau1989approximants}.}
    \label{fig:convergence}
\end{figure}

Once an estimate of the ground state energy for the infinite patch is available, one can analyze the impact of the boundary conditions on the convergence of the result. It is instructive to compare our results with those obtained in a more traditional approach based on the so-called periodic ``approximants'' of the quasiperiodic pattern and the periodic boundary conditions. As can be seen from Figure \ref{fig:convergence}, the latter yields qualitatively much poorer convergence than the local mirror boundary conditions. Actually, for periodic approximants, the error in the determination of the position of the bottom of the spectrum scales roughly as the inverse of the number of sites, and therefore this error remains always of the same order magnitude as the average spacing between energy levels. This indicates that the periodic boundary conditions deeply perturb the structure of the ground state. In the same time, the fast convergence of $E_0$ with the local mirror boundary augurs well for the preservation of the structure of ground state by these conditions. 
\begin{table}
\caption{\label{tab:penrose}The parameters of the ground state of Penrose tiling. See the caption of Table \ref{tab:ammann} for more details.} 
\begin{indented}
\lineup
\item[]\begin{tabular}{@{}*{3}{l}}
\br                              
Number of sites & $-E_0$ & $\lambda$\\
\mr
18 & 4.23305343333938 & 1.20521878491068\\
39 & 4.23437933689518 & 0.92269656399268\\
90 & 4.23464036852845 & 1.12716797696325\\
217 & 4.23467840010248 & 1.05358366136032\\
539 & 4.23468457004802 & 1.08431969726415\\
1365 & 4.23468541949373 & 1.07161692503250\\
3500 & 4.23468555355159 & 1.07645465984201\\
9045 & 4.23468557177912 & 1.07449624769043\\
23490 & 4.23468557463889 & 1.07521692440751\\
61191 & 4.23468557502652 & 1.07492814309309\\
159705 & 4.23468557508806 & 1.07503362860260\\
$\infty$ & 4.2346855750975(2) & 1.07500(1)\\
\br
\end{tabular}
\end{indented}
\end{table}
\par
\begin{figure}[!h]
  \centering
    \includegraphics[width=0.8\textwidth]{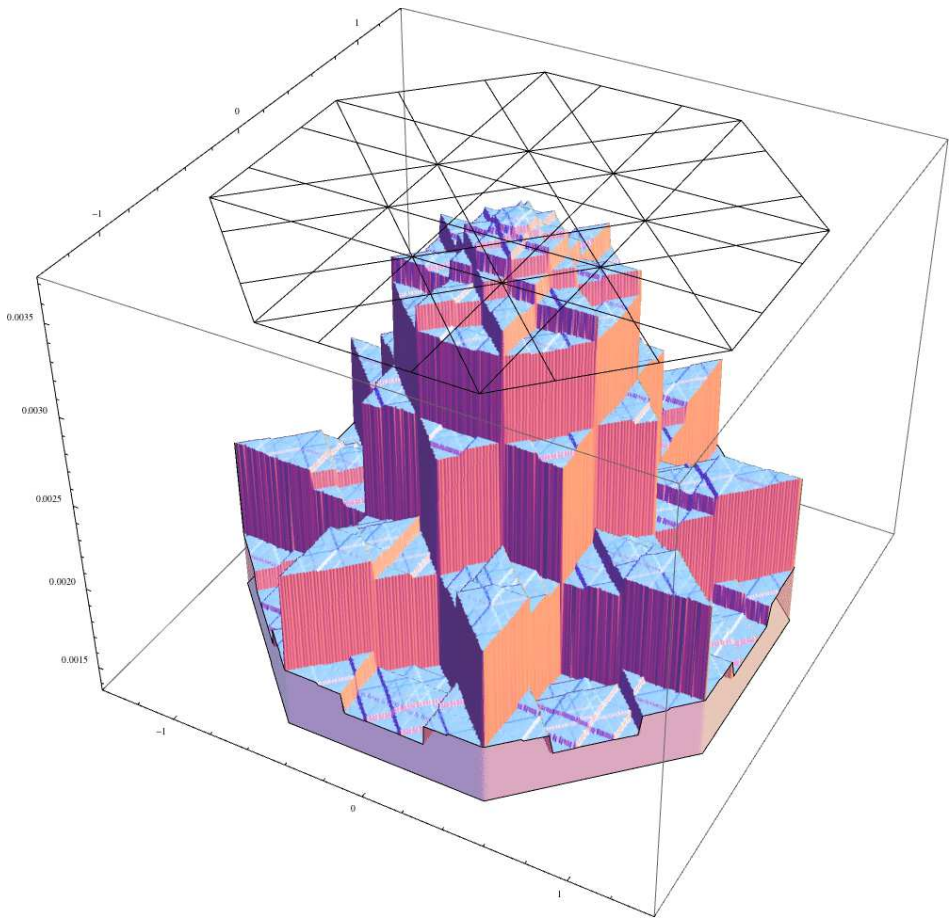}
    \caption{The pre-exponential factor $\Psi_0$ of (\ref{ansatz-tb}) for the ground state on Ammann-Beenker tiling. The superimposed drawing depicts the space $D_n$ (see (\ref{dn})) for $n=2$.}
    \label{fig:cake}
\end{figure}

Since the value of $\lambda$ determines completely the exponential factor $\exp\left(2 \pi \rmi f_{\Lambda}(p)\right)$ in (\ref{ansatz-tb}), the estimation of the former allows one to explore numerically the prefactor $\Psi_0$. Figure \ref{fig:cake} shows the plot of $\Psi_0$ for the ground state of the patch of 138601 sites as a function on the canonical transversal $\Xi$. At first glance, the function does not look like continuous, and it is indeed not so in the coarse topology of the ``atomic surface'' (here an octagon). Note however that the visible step-like discontinuities of $\Psi_0$ are aligned along the boundaries of the spaces $D_n$ (see Figure \ref{fig:c0}, for illustration we also superimposed a representation of $D_2$ on Figure \ref{fig:cake}). Actually, for any given $n$ the function $\Psi_0$ is still discontinuous on $D_n$, but the amplitude of discontinuities visibly decreases with increasing $n$. This is exactly the behaviour one would expect for a function continuous in the Cantor set topology of $\Xi$.
\par
In order to characterize $\Psi_0$ in a more quantitative way, it is convenient to introduce an alternative measure for the regularity of functions on $\Xi$. Given a vector  $v \in L^2(\Xi, \mu_{\rm T})$, let us define its squared residual norm $R_n(v)$ as
\begin{equation}
\label{residual}
R_n(v) = \|\left(\mathcal{P}_n -1\right)v\|^2,
\end{equation}
where $\mathcal{P}_n$ is the orthogonal projection on the subspace $\mathcal{W}_n \subset L^2(\Xi, \mu_{\rm T})$. The rate of decrease of $R_n(v)$ with respect to $n$ characterizes the regularity of $v$ in the following sense. Let $\mathcal{H}(r)$ be the weighted Hilbert space introduced in Section \ref{sec:tight-binding} with the weights $r_n=n^\beta$ for some real $\beta>0$. We shall show that if $R_n(v)=\mathrm{O}(n^{-\alpha})$ for some $\alpha>\beta$, then $v \in \mathcal{H}(r)$. Consider the following formal infinite sum:
\begin{equation}
\label{sum}
S=\sum_{n=1}^\infty \sum_{k=0}^n 
\left((k+1)^\beta-k^\beta)\right)\left(R_n(v)-R_{n+1}(v)\right).
\end{equation}
Since $R_n(v)\ge R_{n+1}(v)$, the terms of (\ref{sum}) are non-negative, and one can interchange the order of summation without affecting convergence:
$$
S=\sum_{k=0}^\infty \left((k+1)^\beta -
k^\beta\right)R_{k+1}(v).
$$
This series converges since $R_{n+1}(v) \left((n+1)^\beta - n^\beta\right)= \mathrm{O}(n^{\beta-\alpha-1})$, and therefore the series (\ref{sum}) converges as well. On the other hand, the summation over $k$ in (\ref{sum}) yields
$$
S=\sum_{n=1}^\infty n^\beta\left(R_n(v)-R_{n+1}(v)\right)= \|v\|_r.
$$
Therefore $\|v\|_r<\infty$ and $v \in \mathcal{H}(r)$.
\par
\begin{figure}[!h]
        \begin{center}
        \subfigure[]{\label{fig:err_a}\includegraphics[width=0.45\textwidth]{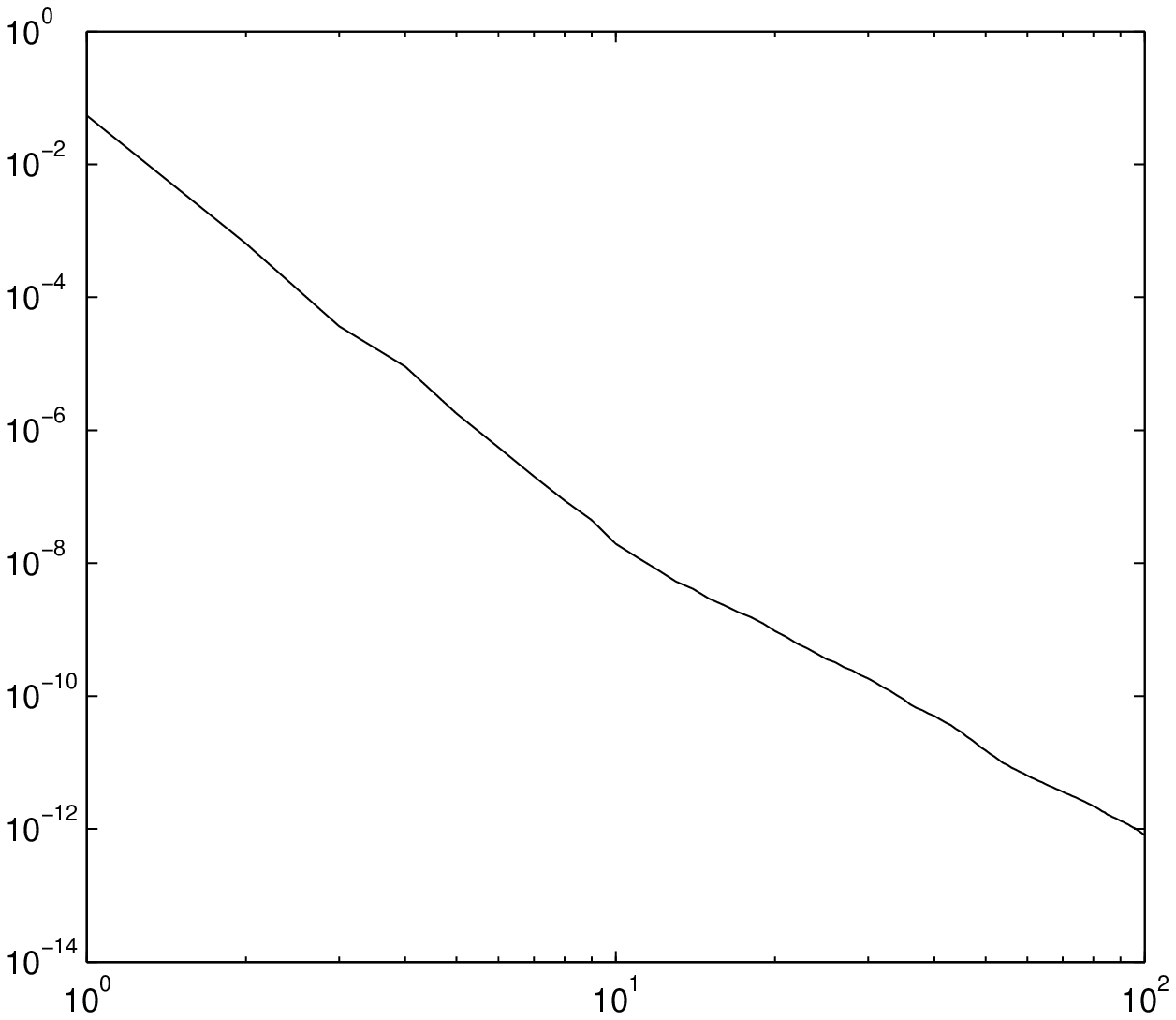}}
        \subfigure[]{\label{fig:err_p}\includegraphics[width=0.45\textwidth]{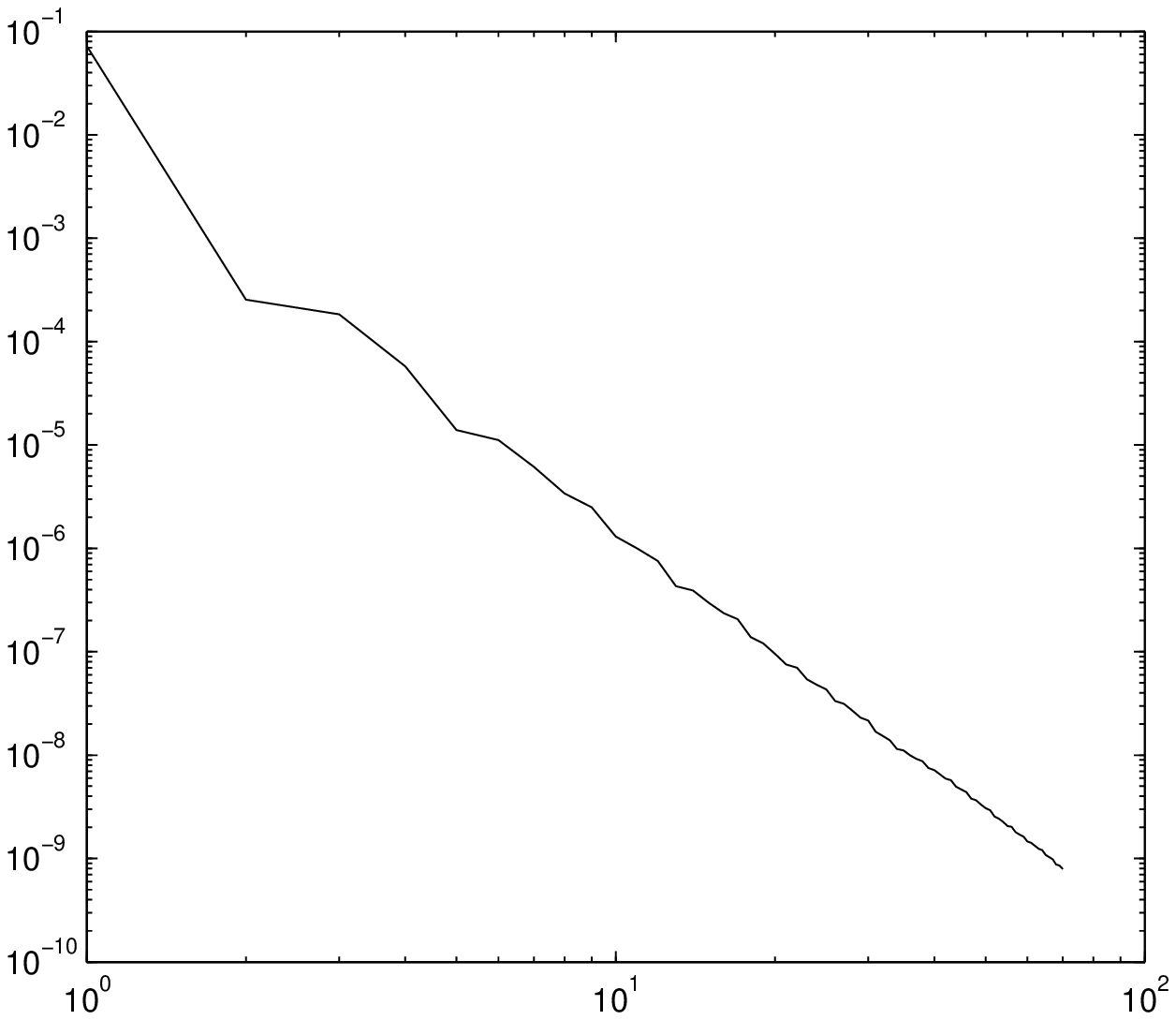}}
        \end{center}
        \caption{The squared residual norms (\ref{residual}) $R_n(\Psi_0)$ for the ground state on Ammann-Beenker \subref{fig:err_a} and Penrose \subref{fig:err_p} tilings as functions of $n$.}\label{fig:residual} 
\end{figure}

For a continuous function on $\Xi$, its integral with the translation-invariant measure $\mu_{\rm T}$ can be approximated by the average of the values of this function at the points from the finite set $\mu(P)$, where $P$ is the set of vertices of a large finite patch of the tiling. This allows for numerical estimation of the squared residual norm (\ref{residual}) of $\Psi_0$. The results for the ground state of Ammann-Beenker and Penrose tilings are shown on Figure \ref{fig:residual}. The plots suggest the power-law decay $R_n(\Psi_0) \sim n^{-\alpha}$ with the exponent $\alpha \approx 4.2$ for Ammann-Beenker and $\alpha \approx 3.9$ for Penrose tiling. Since for any $\beta<\alpha$ and the weights $r_n=n^\beta$ one has $\Psi_0\in \mathcal{H}(r)$, the figures above are comfortably beyond the continuity threshold $\beta=3$ given by (\ref{r_power_law}). 
\par
Notice that up to now, the metric space structure of $\Omega$ was mostly considered as auxiliary, needed only to provide the hull with the topology of a compact Hausdorff space. However, it may play a more important role than it was previously thought. Indeed, Figure \ref{fig:residual} exhibit a fairly regular power-law decay of the squared residual norm of $\Psi_0$. This suggests that the pre-exponential factor $\Psi_0$ is not only continuous on the the hull (or on its canonical transversal in the case of tight-binding model), but also belongs to a more restricted H\"older class. This also indicates that the weighted Hilbert spaces $\mathcal{H}_r$ with power-law weights $r_n=n^\alpha$ may be a natural framework for $\Psi_0$.
\subsection{Stability of results}
The results of the previous section argue strongly in favor of the hypothesis that the ground state wave function in quasicrystals has the form (\ref{ansatz}) (or (\ref{ansatz-tb}) for tight-binding models). Indeed, unlike \cite{sutherland}, we did not start with a predefined wave function and later adjust the Hamiltonian accordingly, but instead analyzed the ground state of two quite standard models. However, the result still may be a matter of mere coincidence. To rule out this possibility, we studied the effect of a continuous deformation on both models.
\par
\begin{figure}[!h]
        \begin{center}
        \subfigure[]{\label{fig:err_diag_a}\includegraphics[width=0.45\textwidth]{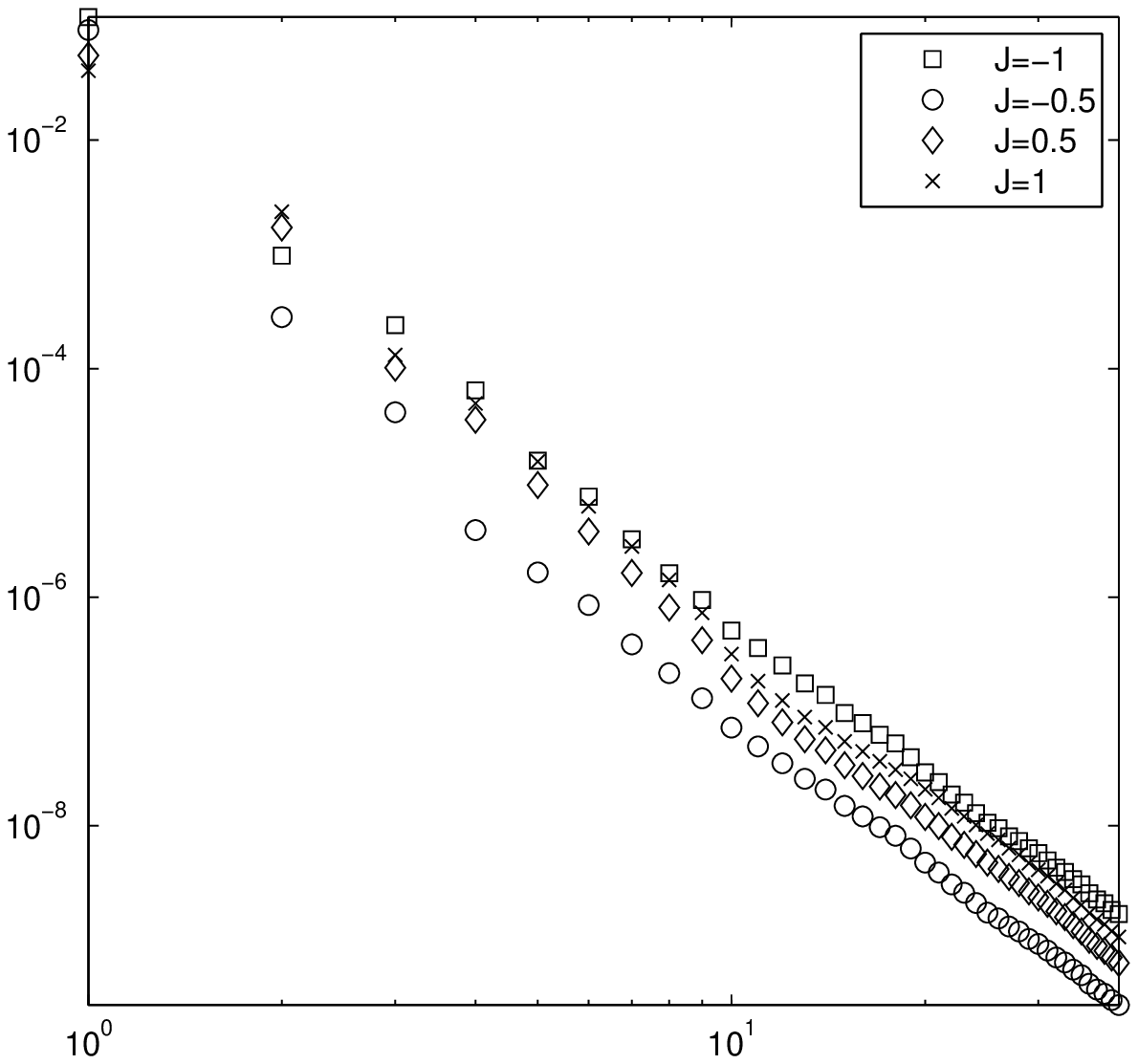}}
        \subfigure[]{\label{fig:err_diag_p}\includegraphics[width=0.45\textwidth]{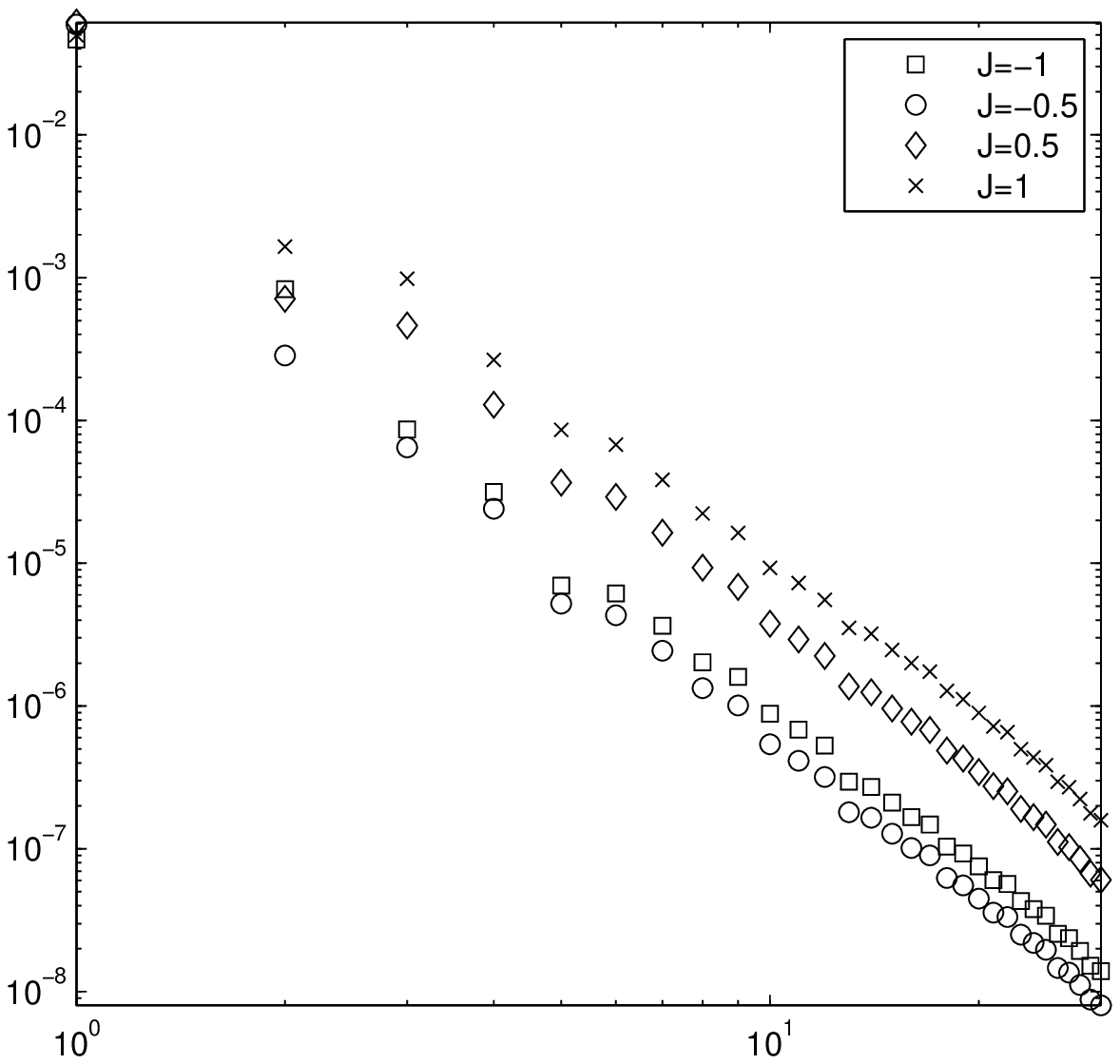}}
        \end{center}
        \caption{The squared residual norms (\ref{residual}) $R_n(\Psi_0)$ for the ground state on deformed Ammann-Beenker \subref{fig:err_diag_a} and Penrose \subref{fig:err_diag_p} tilings as functions of $n$ for different values of the diagonal hopping amplitude $J$.}\label{fig:residual_diag} 
\end{figure}

The choice of the deformation is determined by two considerations. First of all, one has to preserve a particular form of the pattern-equivariant cocycle used to construct the function $f_{\Lambda}$ in (\ref{ansatz-tb}). This can be achieved by choosing the deformation within the {\em mutual local derivability} (MLD) class of the original tiling \cite{baake1995geometric}. The second criterion is less essential since it is due to the limitation of our approach to the mirror boundary conditions. Namely, the formula (\ref{neumann}) is obtained in the assumption that none of the tiling edges crosses the local mirror symmetry lines. These two considerations limit the possible deformations to adding an on-site energy to the Hamiltonian, depending of the local environment of the site, and to introducing the second neighbour hopping along the diagonals of rhombi, moreover for the Penrose tiling only short diagonals of thin rhombi and the long diagonals of thick rhombi are allowed. We have tested these models for several values of the diagonal hopping amplitude $J$ (note that for the Penrose tiling only the short diagonals of thin rhombi were added). The results shown on Figure \ref{fig:residual_diag} suggest that the deformation of the model does not affect the power-law decay of the squared residual norm $R_n(\Psi_0) \sim n^{-\alpha}$ and has no visible effect on the exponent $\alpha$.
\section{Discussion}
The ansatz (\ref{ansatz}) and (\ref{ansatz-tb}) for the single-particle wave function in quasicrystals is universal in that its parameterization by $\Lambda \in \check{H}^1(\Omega)$ depends only on the topology of the hull $\Omega$ of the structure and also in that it covers both continuous and tight-binding models. Numerical evidence suggests that for various tight-binding models on quasiperiodic tilings the the ground state actually has the form (\ref{ansatz-tb}). This makes the proposed wave function a serious contender for the the general form of an eigenstate of Schr\"odinger operator in quasicrystalline potential. 
\par
Similar hierarchical generalized eigenstates have been proposed earlier in the literature, but as far as we know, they all can be described in the framework of our ansatz. In particular, we have already seen that the ``potential'' of Sutherland \cite{sutherland} is proportional to our $f_\Lambda$ for Penrose tiling, while the pre-exponential factor (corresponding to our $\Psi_0$) is merely 1. It is worth mentioning here two generalizations of the approach of Sutherland proposed for Penrose tiling in \cite{repetowicz1998exact}. In the first one, the authors introduced the dependency of the pre-exponential factor on the type of the vertex, which corresponds to a non-constant $\Psi_0$, belonging to a subspace of our $\mathcal{W}_2 \subset \mathcal{H}(r)$. The second approach of \cite{repetowicz1998exact} amounts to consider a linear combination of several deflation-scaled ``potentials''. However the deflation, as can be seen from its action on the ``worm'' (Figure \ref{fig:worm}) acts on $\Lambda_{\rm S}$ by merely changing its sign. Therefore, this solution is also described by some effective class $\Lambda_{\rm S} \in \check{H}^1(\Omega)$, with all other parameters absorbed by $\Psi_0$, which now depends on the local environments of finite range, hence $\Psi_0 \in \mathcal{W}_n$ for some finite $n$.
\par
The conservation of the factors $\lambda_i$  (\ref{floquet}) by the time evolution raises the question of integrability of the quantum single-particle problem in quasicrystalline potentials. It is worth noting that there is a variety of ways to define the quantum integrability (see \cite{caux2011remarks} for a review). In this context, by drawing an analogy with crystals, it is natural to call the system integrable if there are enough integrals of motion (the observables commuting with the Hamiltonian) for having a joint spectrum of finite multiplicity. In this sense, the case of periodic potentials is clearly integrable (the joint spectrum of the Hamiltonian and the components of the quasimomentum is the graph of energy bands). The Fibonacci chain is also integrable in this interpretation since in general case the spectral multiplicity of one-dimensional Schr\"odinger Hamiltonians equals 2 \cite{gilbert1989subordinacy}. By analogy with quasimomentum, it would be natural to consider a normal operator $\hat \lambda_i$, commuting with the Hamiltonian, such that $\hat \lambda_i \psi_\Lambda =\exp\left(2 \pi \rmi \Lambda(c_i)\right) \psi_\Lambda$. Unfortunately, there is no obvious way to do this since the functions $\psi_\Lambda$ are not readily usable for the decomposition of the unity operator. However, it is still possible to address the problem numerically. One feasible experiment would be to study two-dimensional tight-binding models with Cantor set spectrum, for instance the model considered in \cite{benza1991band}. Since the behaviour of the states corresponding to the upper edges of the energy gaps should be similar to that of the ground state (each of these states is in fact a ground state of an appropriate spectral projection of the Hamiltonian), there are good chances that the boundary conditions described in Section \ref{sec:numerical} will not perturb them significantly. In that event, the set of pairs $(E_0, \lambda)$ for each of these states would approximate the joint spectrum of the Hamiltonian and the hypothetical operator $\hat \lambda$.
\par
In this connection it should be mentioned a different way to assess the integrability numerically, namely by studying of the energy level statistics in finite systems. Indeed, the level repulsion is traditionally interpreted as a hallmark of quantum chaos \cite{haake2010quantum}. It was reported \cite{zhong1998level} that the distribution of level spacing in Ammann-Beenker tiling is consistent with that of a random matrix from the Gaussian orthogonal ensemble. However, one should bear in mind that the conditions on the boundary of the studied finite patch of the tiling may affect the level distribution significantly. Indeed, as can be seen from Figure \ref{fig:convergence}, even the periodic conditions perturb the ground state strongly enough to cause an energy shift of the order of the average level spacing. Although the boundary conditions described in Section \ref{sec:boundary} respect the structure of the ground state, we are not aware of any way to do so for all states of a finite patch. As long as this question remains unanswered, making judgements on integrability of Hamiltonians with quasicrystalline potentials based on the level statistics seems premature.  
\par 
It would also be of interest to clarify the role of the matching rules in our results. Let us consider, for instance, the potential of the Fibonacci chain. In this case, $\Omega$ can also be approximated by a sequence of CW-spaces $X_n$ even though no matching rules exist for the Fibonacci chain (as for no other one-dimensional quasiperiodic sequence for that matter). The ``forbidden space'' $\A$ for the Fibonacci chain is a straight segment parallel to $E$ and the homotopy type of $X_n$ is that of the bouquet of two circles. However, the generalized eigenstates in the Fibonacci chain are not described by the ansatz. Indeed, let us consider the solutions of the stationary Schr\"odinger equation. They correspond to sections of a locally transversally constant sheaf on $X_n$, which is clearly a local system of dimension 2. The monodromy of this sheaf with respect to the generators of the fundamental group of $X_n$ is given by the transfer matrices of two basic intervals of the Fibonacci chain. Thus, the monodromy of this sheaf with respect to the cycle encircling the ``forbidden set'' corresponds to the multiplicative commutator of the transfer matrices, which does not commute with the transfer matrices themselves. Therefore, the eigenstates in the Fibonacci chain cannot be described by the formula (\ref{ansatz}), since in the latter the monodromy is given by the scalar factor (\ref{floquet}). The same is true for direct product of $d$ Fibonacci chains and other separable $d\mbox{-dimensional}$ quasiperiodic potentials, none of which possess matching rules. However, the exact role of the matching rules for the ansatz to hold is still unclear.
\par
The above considerations also open the possibility to extend the ansatz (\ref{ansatz}) to the case of non-commutative monodromy. In fact, contrarily to the hull itself, its shape approximants $X_n$ can also be characterized by such homotopy invariant as the fundamental group $\pi_1(X_n)$. This group is non commutative as long as the the ``forbidden space'' $\A$ contains at least three affine subtori of codimension 2 intersecting at one point. This is the case for all models with matching rules.
\par
Finally, let us discuss possible experimentally detectable implications of the hypothesis that the ansatz (\ref{ansatz}) provides a full system of eigenstates for the Schr\"odinger operator in quasicrystals. We have seen that numerical results for the ground states in Amman-Beenker and Penrose tilings yield purely real values for the factor $\lambda_i$ in (\ref{floquet}). This is rather unexpected since this factor plays the role similar to that of Bloch-Floquet multiplier. In a sense, we deal here with an evanescent wave, although this wave propagates along the ``direction'' $f_{\Lambda_{\rm S}}$.  As follows from (\ref{loggrowth}), the span of the structure in this ``direction'' is only logarithmic in its real size; this could explain the power-law behaviour of the occupation rate of the eigenstates in quasicrystals. Let us assume that $\mathrm{coker}(\alpha_n^*)$ is has dimension 1, that is there is only one (up to a factor) cohomology class $\Lambda_{\rm S}$ (for 3D models this is the case for Danzer and canonical $D_6$ tilings). Then we would expect anomalously small conductivity between the points corresponding to the global extrema of the corresponding function $f_{\Lambda_{\rm S}}$. In fact, if an eigenstate is large at the minimum of $f_{\Lambda_{\rm S}}$, it must be small at the maximum and vice versa. Therefore, the global minimum and maximum of $f_{\Lambda_{\rm S}}$ should behave as ``mutually blind spots'' for the propagation of electrons. 
\ack
The authors are grateful to J~Kellendonk for stimulating discussions.
\section*{References}
\bibliography{floquet_qc_iop}
\end{document}